\documentclass[12pt, journal, draftclsnofoot, onecolumn]{IEEEtran}

\usepackage{graphicx}

\usepackage{amsmath}
\usepackage{amsfonts}
\usepackage{amssymb}
\usepackage{psfrag}
\usepackage{cite}
\usepackage{xcolor}
\usepackage{multirow}
\usepackage{subfigure}
\usepackage[linesnumbered,lined,commentsnumbered]{algorithm2e}
\usepackage{hyperref}
\newtheorem{theorem}{\underline{Theorem}}
\newtheorem{lemma}{\underline{Lemma}}
\newtheorem{corollary}{\underline{Corollary}}

\newtheorem{remark}{\underline{Remark}}

\newtheorem{problem}{\underline{Problem}}
\newtheorem{example}{\underline{Example}}
\newtheorem{property}{\underline{Property}}

\newtheorem{proposition}{\underline{Proposition}}
\newtheorem{definition}{\underline{Definition}}

\newcommand{\QED}{{\rm $\blacksquare$}}

\begin{document}
\title{Uplink Non-Orthogonal Multiple Access with Finite-Alphabet Inputs}
{\author{Zheng Dong, He Chen, Jian-Kang Zhang, Lei Huang, and Branka Vucetic
\thanks{Z. Dong is  with Shenzhen University, China and he is also with McMaster University, Canada (email: dongz3@mcmaster.ca), H. Chen and B. Vucetic are with The University of Sydney, Australia (email: \{he.chen, branka.vucetic\}@sydney.edu.au), J.-K. Zhang is with McMaster University, Canada (email: jkzhang@mail.ece.mcmaster.ca), Canada, and L. Huang is with Shenzhen University, China (email: lhuang8sasp@hotmail.com).
}}
\maketitle

\begin{abstract}
This paper focuses on the non-orthogonal multiple access (NOMA) design for a classical two-user multiple access channel (MAC) with finite-alphabet inputs.
In contrast to most of existing NOMA designs using continuous Gaussian input distributions, we consider practical quadrature amplitude modulation (QAM) constellations at both transmitters, the sizes of which are assumed to be not necessarily identical. We propose to maximize the minimum Euclidean distance of the received sum-constellation with a maximum likelihood (ML) detector by adjusting the scaling factors (i.e., instantaneous transmitted powers and phases) of both users. The formulated problem is a \emph{mixed continuous-discrete} optimization problem, which is nontrivial to resolve in general. By carefully observing the structure {of the objective function}, we discover that Farey sequence can be applied to {tackle} the formulated problem. However, the existing Farey sequence is not applicable when the constellation sizes of the two users are not the same. Motivated by this, we define a new type of Farey sequence, termed \emph{punched Farey sequence}. Based on this, we manage to achieve {a closed-form} optimal solution to the original problem by first dividing the entire feasible region into a finite number of Farey intervals and then taking the maximum over all the possible intervals. The resulting sum-constellation is proved to be a regular QAM constellation of a larger size, and hence a simple quantization receiver can be implemented as the ML detector for the demodulation. Moreover, the superiority of NOMA over time-division multiple access (TDMA) in terms of minimum Euclidean distance is rigorously proved. Furthermore, the optimal rate allocation among the two users is obtained in closed-form to further maximize the obtained minimum Euclidean distance of the received signal subject to a total rate constraint. An asymptotic solution is also derived to reveal more insights on how to allocate the rate to each user. Finally, simulation results are provided to verify our theoretical analysis and demonstrate the merits of the proposed NOMA over existing orthogonal and non-orthogonal designs.


\end{abstract}
\section{Introduction}
The forthcoming fifth generation (5G) cellular systems are envisioned to support three generic services, including extreme mobile broadband (eMBB), massive machine-type communications (mMTC), and ultra-reliable and low-latency communications (uRLLC)~\cite{Popovski16ComMag, Lien17ComMag}. 
These diverse services, driven by the explosive growth of mobile data traffic and expected wide roll-out of Internet of Things (IoT), pose challenging requirements for the air interface of wireless networks  where enhanced multiple access technologies are essential.
Apart from several other potential technologies such as massive multiple-input multiple-output (MIMO) and millimeter wave (mmWave) communications, non-orthogonal multiple access (NOMA) has recently emerged as a key enabling radio access technology 
to meet these unprecedented requirements of 5G networks, due to its inherent advantages of high spectral efficiency, massive connectivity, and low transmission latency~\cite{Dai15, Zhiguoding15, Ding2017jsacsurvey, liping06vtmag, qianli14}. 
{\color{black}The concept of NOMA has multiple {variants}, such as power-domain NOMA, sparse code multiple access, pattern division multiple access, low density spreading, and lattice partition multiple access~\cite{Ding2017jsacsurvey}. 
In this paper, we mainly consider the power-domain NOMA.}

The basic principle of NOMA is to serve more than one user with distinct channel conditions simultaneously in the same orthogonal resource block along the time, frequency, or code axes. This can be achieved by applying the superposition coding (SC) at the transmitter as well as multiuser detector (e.g., successive interference cancellation (SIC)) at the receiver side to distinguish the co-channel users.
As such, NOMA is fundamentally different from conventional orthogonal multiple access (OMA) methods primarily used in the previous generations of mobile systems, where each user is allocated to one dedicated orthogonal radio resource block exclusively. {This in turn means that}, in OMA, multiuser communication can be decomposed into several parallel single-user ones free of inter-user interference, and then the well-established single-user encoding/decoding methods can be directly applied with a reasonable tradeoff between network throughput and implementation complexity~\cite[Ch.\,14]{goldsmith05}.

Although the OMA schemes have been {widely used in the past several decades}, they generally cannot achieve the whole multiuser capacity region and thus tend to have a lower spectral efficiency than NOMA {approaches}~\cite{Dai15, Ding2017jsacsurvey, Cover06, Kim12}. For example, in OMA, a resource block allocated to a user with a poor channel condition cannot be reused by another user with a much stronger channel state. Apart from that, OMA is in general not scalable. This is because the amount of resource blocks as well as the granularity of user scheduling strictly limit the number of users that can be supported at the same time. On the contrary, by breaking the orthogonality of the radio resource allocation, 
{\color{black}NOMA has been shown to be able to provide better user fairness and improve physical layer security in addition to the advantages mentioned above~\cite{Dai15, Ding2017jsacsurvey}.}


\subsection{Related Work}
{Despite the fact that} the deployment of NOMA as a new radio access technology in next-generation mobile systems is relatively new, the performance of NOMA has been studied extensively in the information theory society for various channel topologies such as broadcast channel (BC)~\cite{Cover72, Bergmans73, Cover98, Caire03}, multiple access channel (MAC)~\cite{Slepian73, Cheng93,Tse98mac,Tse98macpart2}, and interference channel (IC)~\cite{Carleial75, Carleial78, Kobayashi81, Wyner74, Meulen1977}.
However, these results concentrated mainly on the study of the channel capacity region with the assumption of unlimited encoding/decoding complexity, and therefore lie mostly in the theoretical aspects due to their extremely high implementation cost. Thanks to the rapid progress of the radio frequency (RF) chain and the processing capability of mobile devices in the past decades, the implementation of NOMA is becoming more and more feasible and thus has drawn tremendous attention from both academia and industry very recently~\cite{Ding2017jsacsurvey}. 
More specifically, by taking practical constraints on user fairness and/or radio resource management into consideration, NOMA has been investigated in various wireless systems, such as cognitive radio~\cite{Ding16tvt,Yuan16tvt}, cooperative communications~\cite{Ding15clrelay, Ding16wclrelay}, cellular uplink~\cite{Ding14spl, Zhang16uplink}, cellular downlink~\cite{Saito13, Saito14, Benjebbovu13, Ding16accessiot, dong16jstsp}, and multi-cell networks~\cite{Vaezi16, dong2017jsac}. In fact, a two-user downlink scenario of NOMA, known as multiuser superposition transmission (MUST), has already been incorporated in the 3rd Generation Partnership Project (3GPP) Long Term Evolution-Advanced (LTE-A)~\cite{3gpp2015,Lee16msut}.

We note that, up to now, the vast majority of existing NOMA designs assumed the use of Gaussian input signals~\cite{Cover72, Bergmans73, Cover98, Caire03, Slepian73, Cheng93, Carleial75, Carleial78, Kobayashi81,Wyner74, Meulen1977, liping06vtmag, Ding16tvt,Yuan16tvt, Ding15clrelay, Ding16wclrelay,Ding14spl, Zhang16uplink, Saito13, Saito14, Benjebbovu13, Ding16accessiot,Vaezi16}. Although the Gaussian input is of great significance both theoretically and practically, its implementation in reality will require huge storage capacity, unaffordable computational complexity and extremely long decoding delay~\cite[Ch.\,9]{Cover06}. More importantly, the actual transmitted signals in real communication systems are drawn from finite-alphabet constellations, {such as pulse amplitude modulation (PAM), quadrature amplitude modulation (QAM), and phase-shift keying (PSK)\,\cite[Ch.\,5]{goldsmith05}}. Applying the results derived from the Gaussian inputs to the signals with finite-alphabet inputs can lead to significant performance loss~\cite{verduit06}. In this sense, Gaussian input serves mostly as the theoretical benchmark.

Motivated by the above facts, the NOMA design with finite-alphabet inputs is of utmost importance and has attracted considerable efforts, see e.g.,~\cite{Harshan11, Rajan13, Xiaotc15, Shieh2016tcom, Dytso15, dong2017jsac} and references therein. 
The main principle\footnote{Note that the principle was originally proposed in the seminal work~\cite{Shu76,Shu78,Ahlswede99}, wherein the finite-length codeword design problem in the binary domain were considered from an information-theoretical perspective. } of these {efforts} is to ensure that the signal originated from each user can be uniquely decoded from the received sum-signal at the receiver side. 
By using mutual information as a performance measure, references \cite{Harshan11, Rajan13} considered the NOMA design in {an ideal} two-user Gaussian MAC with finite-input constellations under individual power constraint on each user. Specifically, NOMA was realized by strategically introducing certain constellation rotations (CR) to the adopted PSK signals in~\cite{Harshan11} or using proper power control in~\cite{Rajan13}. However, only numerical solutions to the optimal NOMA designs were provided in \cite{Harshan11, Rajan13}. Moreover, linear precoders were considered for the MIMO MAC in~\cite{Xiaotc15}, where the expression of the weighted sum-rate was asymptotic and the optimal solution was also numerical. Besides, the downlink NOMA system with discrete input distributions was studied in~\cite{Shieh2016tcom}, where the solution is intuitive based on the deterministic approximation of the actual fading channel. The discrete input alphabets were also considered for a two-user interference channel to evaluate the capacity inner bound in~\cite{Dytso15}. In other words, all NOMA designs provided in~\cite{Harshan11, Rajan13, Xiaotc15, Shieh2016tcom, Dytso15} used mutual information as the performance measure, where the solutions were numerical and limited insights on the relationship between the sum-constellation and each user's constellation can thus be drawn from the obtained solutions. 

\subsection{Motivation and Contributions}
Inspired by the aforementioned work, in this paper we target a \emph{closed-form} NOMA design for a classical two-user Gaussian MAC with finite-alphabet inputs {and an optimal maximum likelihood (ML) detector at the receiver}, where the two users are allowed to transmit simultaneously in the same frequency band.
Finding the capacity bound of a Gaussian MAC with Gaussian inputs and adaptive power control has always been {a classic problem}, see e.g.,~\cite{Slepian73, Wyner74, Meulen1977, Cheng93,Tse98mac,Tse98macpart2, Imari14uplink, Zhang16uplink}; the optimal power control scheme for the Gaussian MAC with finite-alphabet inputs, however, is still an open problem and only numerical solutions are available~\cite{Harshan11,Rajan13, Lee16msut, Viterbo16ctw}. 
To fill this gap, in this paper we, for the first time, investigate the optimal power control problem for the two-user Gaussian MAC with \emph{finite square QAM constellations} that maximizes the minimum Euclidean distance of the received signals with the maximum likelihood (ML) detector. Note that QAM signaling is more spectrally efficient than other commonly-used constellations such as PSK signaling. 
Nevertheless, the NOMA design with QAM is more challenging than that with PSK since in QAM both the amplitude and the phase of the modulated signal vary, while in PSK only the phase is different, and thus the unambiguity of the sum-constellation at the receiver side is much more difficult to maintain. {\color{black} Here, it is worth pointing out that in our previous work~\cite{dong2017jsac}, the NOMA design for the Gaussian Z-channel with QAM constellations was investigated, which incorporates the considered two-user MAC as a special case. {In particular}, to resolve the formulated problem, Farey sequence~\cite{Hardy75} was introduced to characterize the minimum Euclidean distance of the sum-constellation. However, due to the inherent symmetric structure between numerators and denominators of the conventional Farey sequence, our results presented in~\cite{dong2017jsac} refer to the case where both transmitters need to use an \emph{identical} constellation size implying the same transmission rate. However, the transmission rates of the users are \emph{not necessarily the same} in practice due to their distinct quality of service (QoS) requirements. To our best knowledge, the NOMA design in terms of power control at users for the Gaussian MAC with not necessarily identical QAM constellations still remains an open problem.


The main contributions of this paper can be summarized as follows:
\begin{enumerate}
\item  We develop a practical NOMA design for the classical two-user complex Gaussian MAC, where the two users are allowed to adopt not necessarily the same QAM constellations. In our design framework, we aim to maximize the minimum Euclidian distance of the received sum-constellation at the receiver side, which dominates the error performance of the considered system, by adjusting the transmit power {and phase} of each user. To this end, we first decompose the complex MAC design problem into two real MAC design problems by strategically rotating the phase of the input signals at the two users. Nevertheless, the decomposed problems are still non-trivial due to their mixed continuous-and-discrete feature. Furthermore, our Farey sequence-based design framework developed in \cite{dong2017jsac} can no longer be applied here due to the fact that the two users may use different QAM constellations.

\item To address this challenging problem, we define a new type of Farey sequence, termed \emph{punched Farey sequence}, which is essential for our NOMA design with not necessarily the same QAM constellations. This concept is even mathematically new to the best of our knowledge~\cite{Hardy75}. We identify and rigourously prove several important properties of the punched Farey sequence in parallel to the conventional Farey sequence. 
Based on the punched Farey sequence and its important properties, we manage to resolve the above decomposed problem for each channel branch by providing a neat closed-form optimal solution, which reveals that the optimal sum-constellation is a regular QAM constellation of a larger size. Due to this nice structure of the sum-constellation, a simple quantization decoder can be employed to implement the ML detector.

\item Based on the obtained closed-form solution, we prove the superiority of this NOMA design over the time-division multiple access (TDMA) approach in terms of the minimum Euclidean distance at the receiver for arbitrary given channel realization and rate allocation. Actually, this is a surprising result since the new NOMA method can achieve a better error performance than TDMA in a high SNR regime even if there is no near-far effect. Furthermore, we also address the optimal rate-allocation problem among the two users to maximize the minimum Euclidean distance of the received sum-constellation subject to a total rate constraint. More importantly, we derive a high-rate approximate solution to the optimal rate-allocation problem, which uncovers a lot of insights on the practical system designs.
\end{enumerate}	

\section{Two-User Gaussian Multiple-Access Channel}
We consider a two-user Gaussian MAC given by
\begin{align}\label{eqn:comlexMAC}
z &=h_1 x_1+ h_2 x_2 + \xi,
\end{align}
where $z$ is the received signal at the base station (BS), $h_k$ denotes the complex channel coefficient  between the transmitter $S_k$ and BS for $k=1,2$, and $\xi$ is the additive zero-mean, circularly symmetric complex Gaussian (CSCG) noise with variance $2\sigma^2$, i.e., $\xi \sim \mathcal{CN}(0, 2\sigma^2)$.
We assume that perfect channel state information (CSI) is available to all the nodes\footnote{The optimal design can also be performed at the BS which sends the results back to the transmitters via the forward links. In this case, only BS needs to know the  full CSI.} and symbol synchronization is maintained at BS. The transmitted symbols $x_k$ are superimposed at the receiver in a NOMA manner which are chosen randomly, independently and equally likely from the (finite) square QAM constellation $\mathcal{Q}_k$, and are subject to individual average power constraint $P_k$, i.e., $\mathbb E[|x_k|^2]\le P_k$ for $k=1,2$.

Although we use a complex baseband representation in~\eqref{eqn:comlexMAC}, the modulated and demodulated signals are real since the oscillator at the transmitter can only generate real sinusoids rather than complex exponentials, and the channel then introduces amplitude and phase distortion to the transmitted signals~\cite{goldsmith05}. As such, we follow~\cite{Rajan13} to decompose the considered complex Gaussian MAC given in~\eqref{eqn:comlexMAC} into two parallel real-scalar Gaussian MACs, which are called the in-phase and quadrature components, respectively~\cite{goldsmith05}.
This means that the original two-dimensional QAM constellation can be split into two one-dimensional PAM constellations to be transmitted via the in-phase and quadrature branches. Besides, since the in-phase and quadrature components of the sum-constellation are separable, they can be decoded independently at the receiver, thereby reducing the decoding complexity. Mathematically, we notice that~\eqref{eqn:comlexMAC} is equivalent to
\begin{align}\label{eqn:comlexMACrotation}
z &=|h_1| x_1 \exp(j \arg(h_1)) + |h_2| x_2 \exp(j \arg(h_2)) + \xi.
\end{align}
{To simplify the subsequent expressions,} we let $y= {\rm Re}(z)$, $y'={\rm Im}(z)$, $w_1 s_1={\rm Re}(x_1 \exp(j \arg(h_1)))$, $w_1' s_1'={\rm Im}(x_1 \exp(j \arg(h_1)))$, $w_2 s_2={\rm Re} (x_2 \exp(j \arg(h_2)))$, $w_2' s_2'={\rm Im} (x_2 \exp(j \arg(h_2)))$, $n={\rm Re}(\xi)$ and $n'={\rm Im}(\xi)$, where ${\rm Re}(\cdot)$ and ${\rm Im}(\cdot)$ are the real and  imaginary parts of the complex number, respectively.  Besides, $w_1$, $w_2$, $w_1'$, and $w_2'$ are the real non-negative scalars determining the minimum Euclidean distance of the actual transmitted PAM constellation sets, which are referred to as the \emph{weighting coefficients} throughout this paper. Now, the in-phase and quadrature branches of~\eqref{eqn:comlexMAC}  can be reformulated by
\begin{subequations}
	\begin{align}
	y &= |h_1| w_1 s_1 + |h_2| w_2 s_2 + n,\label{eqn:inphasesubchannel}\\
	y'&= |h_1| w_1' s_1' + |h_2| w_2' s_2' + n',
	\end{align}
\end{subequations}
where $n, n' \sim \mathcal{N}(0, \sigma^2)$ are independent and identically distributed (i.i.d.) real additive white Gaussian components since the complex noise term $\xi$ is assumed to be CSCG noise.

Without loss of generality, we assume that $x_1 \exp(j \arg(h_1)) \in \mathcal{Q}_1$ and $x_2 \exp(j \arg(h_2)) \in \mathcal{Q}_2$, where $\mathcal{Q}_1$ and $\mathcal{Q}_2$ are $M_1^2$- and $M_2^2$-ary square QAM constellations ($M_1$ and $M_2$ are both no less than 2 but not necessarily equal to each other), respectively,  given by
$\mathcal{Q}_1 \triangleq \{\pm w_1 (2k-1) \pm w_1' (2\ell-1)j: k, \ell=1,\ldots, M_1/2 \}$ and
$\mathcal{Q}_2 \triangleq  \{\pm w_2 (2k-1) \pm w_2' (2\ell-1)j:  k, \ell=1,\ldots, M_2/2 \}$.
As a result, the information-bearing symbols  $s_1, s_1' \in \mathcal{A}_{M_1}=\{\pm (2k-1)\}_{k=1}^{{M_1}/2}$, sent by $S_1$, and $s_2, s_2' \in\mathcal{A}_{M_2}=\{\pm (2k-1)\}_{k=1}^{{M_2}/2}$, transmitted by $S_2$, are drawn from the standard PAM constellations with equal probability.
We consider that an equal power allocation between two branches is performed to balance the minimum Euclidean distance of the two PAM constellations~\cite[Ch.\,6.1.4]{goldsmith05} and  the transmitted signals over both subchannels should still be subject to average power constraints, i.e., $\mathbb E[w_1^2 |s_1|^2]\le P_1/2$, $\mathbb E[w_1'^2 |s_1'|^2]\le P_1/2$, $\mathbb E[w_2^2|s_2|^2]\le P_2/2$, and $\mathbb E[w_2'^2|s_2'|^2]\le P_2/2$.

An important problem for the considered MAC is, for any given QAM constellation sizes of both messages, how to optimize the values of scaling coefficients $w_1$, $w_2$, $w_1'$ and $w_2'$ to minimize the average error probability at the receiver, subject to the individual average power constraints at both transmitters. As the in-phase and quadrature subchannels are symmetric, if the same algorithm is applied to both branches, we will expect to have $w_1=w_1'$ and $w_2=w_2'$, and we call $\mathcal{Q}_1$ and $\mathcal{Q}_2$ the symmetric square QAM constellations. It is worth mentioning that our framework can be readily extended to un-symmetric signaling~\cite{Vidal16, Jafar10}, i.e., un-equal power allocation between the two branches. By leveraging the decomposable property of the complex Gaussian MAC and the symmetry of the two subchannels, we can simply focus on the design for one of the two real-scalar Gaussian MACs with PAM constellation sets, which will be elaborated in next section\footnote{It should be pointed out that designing two PAM constellations for both subchannels separately is a practical but not necessarily optimal approach. In fact, this approach has been widely adopted in literature, such as in~\cite{Cadambe08,Tse08, Jafar10,Bresler10, Prasad16, Vidal16}. How to design a two-dimensional complex constellation directly for the Gaussian MAC has been left as a future work.}.


\section{The Weighting Coefficients Design for the Real-Scalar Gaussian MAC}
In this section, we consider the constellation design problem, i.e., finding the optimal weighting coefficients  $w_1$ and $w_2$, for the in-phase real-scalar Gaussian MAC. As the two sub-channels are symmetric, the optimal solution to the quadrature component can be obtained in exactly the same way and hence is omitted for brevity.

\subsection{Problem Formulation}
Recall that 
$\mathbb E[w_1^2 |s_1|^2]\le P_1/2$, $\mathbb E[w_2^2|s_2|^2]\le P_2/2$,  and hence  $0< w_1 \le \sqrt{\frac{3 P_1}{2(M_1^2-1)}}$, $0< w_2 \le \sqrt{\frac{3 P_2}{2(M_2^2-1)}}$. For notation simplicity, we set $|\tilde{h}_1|=\sqrt{\frac{3 P_1}{2(M_1^2-1)}}|h_1|$, $|\tilde{h}_2| =\sqrt{\frac{3 P_2}{2(M_2^2-1)}}|h_2|$ and
\begin{align}\label{eqn:changelabel}
\tilde{w}_1=\sqrt{\frac{2(M_1^2-1)}{3 P_1}} w_1, \tilde{w}_2=\sqrt{\frac{2(M_2^2-1)}{3 P_2}} w_2,
\end{align}	
where $0<  \tilde{w}_1 \le 1$ and  $0<  \tilde{w}_2 \le 1$. The received signal in~\eqref{eqn:inphasesubchannel} can thus be re-written as
\begin{align}\label{eqn:realGaussianMAC}
y &= |\tilde{h}_1| \tilde{w}_1 s_1 +|\tilde{h}_2| \tilde{w}_2 s_2 + n.
\end{align}


We assume that a coherent maximum-likelihood~(ML) detector is used by BS to estimate the transmitted signals in a symbol-by-symbol fashion\footnote{Since we perform a symbol-by-symbol detection, the decoding complexity is {at most} $\mathcal{O}(M_1 M_2)$ with $M_1$ and $M_2$ being the PAM constellation size of $s_1$ and $s_2$, respectively. 
}. Mathematically, the estimated signals can be expressed as
\begin{align*}
&(\hat s_1, \hat s_2)=\arg \min_{(s_1, s_2)}~\big|y -(|\tilde{h}_{1}| \tilde{w}_1 s_1 + |\tilde{h}_{2}| \tilde{w}_2 s_2)\big|.
\end{align*}

By applying the nearest neighbour approximation method~\cite[Ch.6.1.4]{goldsmith05}} at high SNRs for ML receiver, the average error rate is dominated by the minimum Euclidean distance of the received constellation points owing to the exponential decaying of the Gaussian distribution. As such, in this paper, we aim to devise the optimal value of $(\tilde{w}_1, \tilde{w}_2)$ (or equivalently constellations $\mathcal{Q}_1$ and $\mathcal{Q}_2$) to maximize the minimum Euclidean distance of constellation points of the received signal. The Euclidean distance between the two received signals $y(s_1,s_2)$ and $y(\tilde{s}_1, \tilde{s}_2)$ at the receiver for $(s_1, s_2)$ and $(\tilde{s}_1, \tilde{s}_2)$ in the noise-free case is given by
	\begin{align}
	&|y(s_1, s_2) -y(\tilde{s}_1, \tilde{s}_2)|=\big||\tilde{h}_1| \tilde{w}_1(s_1 -\tilde{s}_1) -  |\tilde{h}_2| \tilde{w}_2 (\tilde{s}_2-s_2 ) \big|.
	\end{align}
Note that $s_1$, $\tilde{s}_1$, $s_2$ and $\tilde{s}_2$ are all odd numbers, and thus we can let $s_1 -\tilde{s}_1=2 n$ and $\tilde{s}_2-s_2 =2 m$, in which $n \in \mathbb Z_{M_1-1}$ and $m \in \mathbb Z_{M_2-1}$ with {$\mathbb Z_{N}  \triangleq \{0,\pm 1,\cdots, \pm N\}$} denoting the set containing all the possible differences.  Similarly, we also define $\mathbb Z_{(M_1-1,M_2-1)}^2\triangleq \{(a,b):a\in \mathbb Z_{M_1-1}, b\in \mathbb Z_{M_2-1}\}$, and $\mathbb N_{(M_1-1, M_2-1)}^2 \triangleq \{(a,b): a\in \mathbb N_{M_1-1}, b\in \mathbb N_{M_2-1}\}$ where $\mathbb {N}_{N}\triangleq \{0,1,\cdots, N\}$. From the definitions above,  $(s_1, s_2) \neq (\tilde{s}_1, \tilde{s}_2)$ is equivalent to $(m,n) \neq (0,0)$ (i.e., $m\neq 0$ or $n \neq 0$).
To proceed, we define
\begin{align}
d(m,n) &=\frac{1}{2}|y(s_1, s_2) -y(\tilde{s}_1, \tilde{s}_2)|\nonumber\\
       &=\big||\tilde{h}_1|\tilde{w}_1 n -  |\tilde{h}_2| \tilde{w}_2 m\big|,~ (m, n) \in \mathbb Z_{(M_1-1, M_2-1)}^2 \setminus \{(0,0)\},\label{eqn:euclideandist}
\end{align}
{where $\mathcal{A} \setminus \mathcal{B}\triangleq\{x\in \mathcal{A} {\rm~and~} x\notin \mathcal{B}\}$}.
We are {at a point to} formally formulate the following max-min optimization problem,
\begin{problem}\emph{Power Control of NOMA in real-scalar MAC with PAM constellation:}\label{pbm:maxminopt}
Find the optimal value of $(\tilde{w}_1^*, \tilde{w}_2^*)$ subject to the individual average power constraint such that the minimum Euclidean distance $d^*$ of the received signal constellation points is maximized, i.e.,
	\begin{subequations}\label{pbm:optnoma}
	\begin{align}
	(\tilde{w}_1^*, \tilde{w}_2^*)=&\arg \max_{(\tilde{w}_1, \tilde{w}_2)}~ {\min_{ (m,n) \in \mathbb Z_{(M_1-1,M_2-1)}^2 \setminus\{(0,0)\} } d(m,n)}  \label{eqn:maxminoptobj}\\
{\rm s.t.~} & 0< \tilde{w}_1 \le 1 {\rm~and~} 0< \tilde{w}_2 \le 1.
	\end{align}
	\end{subequations}
	\hfill$\blacksquare$
\end{problem}

Note that the inner optimization variable of finding the minimum Euclidean distances is discrete, while the outer one $(\tilde{w}_1, \tilde{w}_2)$ is continuous. In other words, Problem~\ref{pbm:maxminopt} is a \emph{mixed continuous-discrete} optimization problem and it is in general hard to solve. To the best of our knowledge, only numerical solutions to such kind of problems are available in the open literature~\cite{Harshan11,Rajan13, Lee16msut, Viterbo16ctw}. To optimally and systematically solve this problem, we now develop a design framework based on the \emph{Farey sequence}~\cite{Hardy75}, in which the entire feasible region of $(\tilde{w}_1, \tilde{w}_2)$ is divided into a finite number of mutually exclusive sub-regions.
Then, for each sub-region, the formulated optimization problem can be solved optimally with a closed-form solution, and subsequently the overall maximum value of Problem~\ref{pbm:maxminopt} can be attained by taking the maximum value of the objective function among all the possible sub-regions. We first consider the inner optimization problem in \eqref{pbm:optnoma} given by:
\begin{problem}\emph{Finding differential pairs with the minimum Euclidean distance:}\label{pbm:minEuclidiandist}
\begin{align}\label{eqn:mindistD1}
\min_{ (m,n)\in \mathbb Z_{(M_1-1, M_2-1)}^2 \setminus\{(0,0)\} }~d(m,n) = \min_{\quad (m,n)\in \mathbb Z_{(M_1-1, M_2-1)}^2 \setminus\{(0,0)\}} \big||\tilde{h}_1| \tilde{w}_1 n -  |\tilde{h}_2| \tilde{w}_2 m\big|.
\end{align}
\hfill\QED
\end{problem}
We should point out that finding the closed-form solution to the optimal $(m,n)$ for~\eqref{eqn:mindistD1} is not trivial since the solution depends on the values of $|\tilde{h}_1|$ and $|\tilde{h}_2|$, which can span the whole positive real axis. Moreover, the values of $\tilde{w}_1$ and $\tilde{w}_2$ will be optimized later and cannot be determined beforehand. It is worth mentioning here that a similar optimization problem was formulated and resolved for a Gaussian Z channel in our previous work \cite{dong2017jsac}. In \cite{dong2017jsac}, we resorted to the existing Farey sequence to solve the formulated problem. However, due to the inherent symmetric structure between numerators and denominators of the conventional Farey sequence, our results presented in~\cite{dong2017jsac} refers only to the case where both transmitters need to use exactly \emph{identical} constellation size (i.e., the same transmission rate) and thus cannot be applied to the problem in this paper with $M_1$ and $M_2$ not necessarily the same. Motivated by this, in this paper we define a new type of Farey sequence, termed punched Farey sequence.
In the subsequent section, we will introduce the definition and some important properties of the original Farey sequence and the developed punched Farey sequence.

\subsection{Farey Sequence}
The Farey sequence characterizes the relationship between two positive integers and the formal definition is  given as follows:
\begin{definition}\emph{Farey sequence~\cite{Hardy75}:}
The Farey sequence $\mathfrak{F}_K$ is the ascending sequence of irreducible fractions between $0$ and $1$ whose denominators are less than or equal to $K$.
\hfill\QED
\end{definition}	

By the definition, $\mathfrak{F}_K=\big(\frac{b_k}{a_k}\big)_{k=1}^{|\mathfrak{F}_K|}$ is a sequence of fractions $\frac{b_k}{a_k}$ such that $0\le b_k\le a_k\le K$ and $\langle a_k, b_k \rangle =1$  arranged in an increasing order, where $\langle a,b\rangle$ denotes the largest common divider of non-negative integers $a,b$. In addition, $|\mathfrak{F}_K| =1+\sum_{m=1}^K \varphi(m)$ is the cardinality of $\mathfrak{F}_K$ with $\varphi(\cdot)$ being the Euler's totient function~\cite{Hardy75}.  
An example of Farey sequence is given as follows:
\begin{example}
    $\mathfrak{F}_5$ is the ordered sequence
	$\big(\frac{0}{1}, \frac{1}{5}, \frac{1}{4}, \frac{1}{3}, \frac{2}{5},\frac{1}{2}, \frac{3}{5}, \frac{2}{3},  \frac{3}{4}, \frac{4}{5}, \frac{1}{1}\big)$.
\end{example}

It can be observed that each Farey sequence begins with number 0 (fraction $\frac{0}{1}$) and ends with 1 (fraction $\frac{1}{1}$). The series of breakpoints after $\frac{1}{1}$ is the reciprocal version of the Farey sequence. We call the Farey sequence together with its reciprocal version as the \emph{extended Farey sequence} which is formally defined as follows:
\begin{definition}\emph{Extended Farey sequence:}
	The extended Farey sequence $\mathfrak{S}_K$ of order $K$ is the sequence of ascending irreducible fractions, where the maximum value of the numerator and denominator do not exceed $K$.
	\hfill\QED
\end{definition}

From the definition, we have $\mathfrak{S}_K=\big(\frac{b_k}{a_k}\big)_{k=1}^{|\mathfrak{S}_K|}$ with $\langle a_k,b_k\rangle=1$ and  $|\mathfrak{S}_K|=1+2\sum_{m=1}^K \varphi(m)$. We have the following example:
\begin{example}
	$\mathfrak{S}_5$ is the sequence
	$\big(\frac{0}{1}, \frac{1}{5}, \frac{1}{4}, \frac{1}{3}, \frac{2}{5},\frac{1}{2}, \frac{3}{5}, \frac{2}{3},  \frac{3}{4}, \frac{4}{5}, \frac{1}{1},\frac{5}{4},\frac{4}{3},\frac{3}{2},\frac{5}{3},\frac{2}{1},\frac{5}{2},\frac{3}{1},\frac{4}{1},\frac{5}{1},\frac{1}{0}\big)$.
\end{example}

It can be observed that the extended Farey sequence starts with number 0 (fraction $\frac{0}{1}$) and end with $\infty$ (fraction $\frac{1}{0}$).
%
%
We now propose a new definition called \emph{Punched Farey sequence} in number theory as follows.
\begin{definition}\emph{Punched Farey sequence:}\label{def:punchedfarey}
The punched (extended) Farey sequence $\mathfrak{P}_K^L$ 
is the ascending sequence of irreducible fractions whose denominators are no greater than $K$ and numerators are no greater than $L$.
\hfill\QED
\end{definition}
\begin{example}
	$\mathfrak{P}_5^2$ is the ordered sequence
	$\big(\frac{0}{1}, \frac{1}{5}, \frac{1}{4}, \frac{1}{3}, \frac{2}{5},\frac{1}{2},  \frac{2}{3}, \frac{1}{1},\frac{2}{1},\frac{1}{0}\big)$.
\end{example}

From Definition~\ref{def:punchedfarey}, when $L=K$, $\mathfrak{P}_K^K$ degenerates into Farey sequence $\mathfrak{F}_K$, i.e., $\mathfrak{P}_K^K=\mathfrak{F}_K$. We can also observe that each punched Farey sequence begins with number 0 (fraction $\frac{0}{1}$) and ends with $\infty$ (fraction $\frac{1}{0}$).

We now develop some elementary properties of the punched Farey sequence in line with Farey sequences~\cite{Hardy75}.
It is worth pointing out that, although for some properties,  we can find the counterparts in conventional Farey sequences, the extension to the punched Farey sequences is non-trivial and the following results are new.

\begin{property}\label{prop:basicFareyprop}
If $\frac{n_1}{m_1}$ and $\frac{n_2}{m_2}$ are  two adjacent terms {(called Farey pairs)} in $\mathfrak{P}_K^L$ ($\min\,\{K,L\}\ge 2$) such that $\frac{n_1}{m_1}< \frac{n_2}{m_2}$, then,
1) $\frac{n_1 + n_2}{m_1 + m_2} \in \big(\frac{n_1}{m_1}, \frac{n_2}{m_2} \big)$, $\frac{m_1+m_2}{n_1 + n_2} \in \big(\frac{m_2}{n_2}, \frac{m_1}{n_1} \big)$;
2) $m_1 n_2 - m_2 n_1=1$;
3) If $n_1+n_2\le L$, then $m_1+m_2>K$ and if $m_1+m_2\le K$, then $n_1+n_2>L$;
4) $n_1+n_2 \ge 1$ where the equality is attained if and only if $\frac{n_1}{m_1}=\frac{0}{1}$ and $\frac{n_2}{m_2}=\frac{1}{K}$. Likewise, $m_1+m_2 \ge 1$ where the equality is attained if and only if $\frac{n_1}{m_1}=\frac{L}{1}$ and $\frac{n_2}{m_2}=\frac{1}{0}$.
	~\hfill$\blacksquare$

\end{property}

The proof is given in Appendix\ref{appendix:propbasicFarey}.


\begin{property}\label{prop:mediumvalue}
	If $\frac{n_1}{m_1}$, $\frac{n_2}{m_2}$  and $\frac{n_3}{m_3}$ are three consecutive terms in $\mathfrak{P}_K^L$ with $\min\,\{K,L\}\ge 2$ such that $\frac{n_1}{m_1}< \frac{n_2}{m_2}<\frac{n_3}{m_3}$, then $\frac{n_2}{m_2}=\frac{n_1 + n_3}{m_1 + m_3}$.
	~\hfill$\blacksquare$
\end{property}

The proof is provided in Appendix\ref{appendix:mediumvalue}.

\begin{property}\label{prop:inequalitymodified}
Consider $\frac{n_1}{m_1},\frac{n_2}{m_2},\frac{n_3}{m_3},\frac{n_4}{m_4} \in \mathfrak{P}_{K}^L$ with $\min\,\{K,L\}\ge 3$, such that $\frac{n_1}{m_1}<\frac{n_2}{m_2}<\frac{n_3}{m_3}<\frac{n_4}{m_4} $ where $\frac{n_2}{m_2}, \frac{n_3}{m_3}$ are successive in $\mathfrak{P}_{K}^L$, then $\frac{n_1+n_3}{m_1+m_3}\le \frac{n_2}{m_2}$ and $\frac{n_3}{m_3} \le \frac{n_2+n_4}{m_2+m_4}$. ~\hfill$\blacksquare$
\end{property}

The proof is provided in Appendix\ref{appendix:inequalitymodified}.

\subsection{The Minimum Euclidean Distance of the  Constellation Points of the Received Signal}
We are now ready to solve {Problem\,\ref{pbm:minEuclidiandist}} to find the  differential pairs $(m,n)$ having the minimum Euclidean distance. To this end, we first introduce the following preliminary propositions.

\begin{proposition}\label{prop:mindistant}
Let $\mathbb F_{(M_1-1, M_2-1)}^2 =\{(m,n): \frac{n}{m}\in \mathfrak{P}_{M_2-1}^{M_1-1} \}$,
and then
\begin{align*}
\min_{\quad (m,n)\in \mathbb Z_{(M_1-1, M_2-1)}^2 \setminus\{(0,0)\} }~d(m,n)=\min_{\quad (m,n)\in \mathbb F_{(M_1-1, M_2-1)}^2 }~d(m,n).~~~~~~~~~~~~~~~~~~~~~~~~~~~~~~~~\blacksquare
\end{align*}
\end{proposition}

The proof is similar to~\cite[App.-A]{dong2017jsac} and hence is omitted for brevity.


\begin{proposition}\label{prop:basicfareyextended}
Let $\frac{n_1}{m_1}$ and $\frac{n_2}{m_2}$ be two terms of $\mathfrak{P}_{M_2-1}^{M_1-1}$ such that $\frac{n_1}{m_1}< \frac{n_2}{m_2}$. Then, for $\frac{|\tilde{h}_{2}| \tilde{w}_2}{|\tilde{h}_{1}| \tilde{w}_1} \in (\frac{n_1}{m_1}, \frac{n_2}{m_2})$ and $d(m,n) =\big||\tilde{h}_{1}| \tilde{w}_1 n-|\tilde{h}_{2}| \tilde{w}_2 m\big|$, we have	
1) If $\frac{|\tilde{h}_{2}| \tilde{w}_2}{|\tilde{h}_{1}| \tilde{w}_1}= \frac{n_1 +n_2}{m_1 + m_2}$, then $d(m_1, n_1)=d(m_2, n_2)$;
2) If $\frac{|\tilde{h}_{2}| \tilde{w}_2}{|\tilde{h}_{1}| \tilde{w}_1} \in \big(\frac{n_1}{m_1}, \frac{n_1 +n_2}{m_1 + m_2}\big)$, then $d(m_1, n_1) <d(m_2, n_2)$;
3) If  $\frac{|\tilde{h}_{2}| \tilde{w}_2}{|\tilde{h}_{1}| \tilde{w}_1} \in \big(\frac{n_1 +n_2}{m_1 + m_2}, \frac{n_2}{m_2}\big)$, then $d(m_2, n_2) <d(m_1, n_1)$.
\hfill$\blacksquare$
\end{proposition}

The proof can be found in Appendix\ref{appendix:fareydiv}.

\begin{proposition}\label{prop:worstcasemodified}
	For any $\frac{n_1}{m_1},\frac{n_2}{m_2},\frac{n_3}{m_3},\frac{n_4}{m_4} \in \mathfrak{P}_{M_2-1}^{M_1-1}$ with $|\mathfrak{P}_{M_2-1}^{M_1-1}| \ge 4$, such that $\frac{n_1}{m_1}<\frac{n_2}{m_2}<\frac{n_3}{m_3}<\frac{n_4}{m_4} $, and $\frac{n_2}{m_2}, \frac{n_3}{m_3}$ are successive in $\mathfrak{P}_{M_2-1}^{M_1-1}$, we have
	1) If $\frac{|\tilde{h}_{2}| \tilde{w}_2}{|\tilde{h}_{1}| \tilde{w}_1} \in (\frac{n_2}{m_2},\frac{n_2+n_3}{m_2+m_3})$, then $\min_{(m,n) \in \mathbb F_{(M_1-1, M_2-1)}^2}~d(m,n)
		=d(m_2, n_2)
		=|\tilde{h}_{2}|\tilde{w}_2 m_2- |\tilde{h}_{1}| \tilde{w}_1 n_2$;
	2) If  $\frac{|\tilde{h}_{2}| \tilde{w}_2}{|\tilde{h}_{1}| \tilde{w}_1} \in (\frac{n_2+n_3}{m_2+m_3},\frac{n_3}{m_3})$, then $\min_{(m,n) \in \mathbb F_{(M_1-1, M_2-1)}^2}~d(m,n)
		=d(m_3, n_3)
		=|\tilde{h}_{1}| \tilde{w}_1 n_3 - |\tilde{h}_{2}|\tilde{w}_2 m_3$.
	~\hfill$\blacksquare$	
\end{proposition}

The proof is given in Appendix\ref{appendix:worstcase}.

\subsection{Closed-Form Optimal Solution to Problem~\ref{pbm:maxminopt}}	
With the propositions presented in the previous subsection, we now can solve Problem~\ref{pbm:maxminopt} by restricting~$\frac{|\tilde{h}_{2}| \tilde{w}_2}{|\tilde{h}_{1}| \tilde{w}_1}$ into a certain {punched Farey interval determined by the corresponding Farey pair} where a closed-form solution is attainable.
More specifically, we consider the punched Farey sequence given by $\mathfrak{P}_{M_2-1}^{M_1-1}=\big(\frac{b_1}{a_1},\frac{b_2}{a_2},\cdots, \frac{b_{C}}{a_{C}}\big)$,
where $C=|\mathfrak{P}_{M_2-1}^{M_1-1}|$. Now, assume that $\frac{|\tilde{h}_{2}| \tilde{w}_2}{|\tilde{h}_{1}| \tilde{w}_1} \in \big(\frac{b_k}{a_k}, \frac{b_{k+1}}{a_{k+1}}\big)$ where $\big(\frac{b_k}{a_k}, \frac{b_{k+1}}{a_{k+1}}\big)$ is the $k$-th punched Farey interval for $k=1,\ldots, C-1$, and we aim to find the optimal
$(\tilde{w}_1^*(k), \tilde{w}_2^*(k))$ such that
\begin{subequations}\label{eqn:maxminbyinterval}
\begin{align}
&g\Big(\frac{b_k}{a_k}, \frac{b_{k+1}}{a_{k+1}}\Big) = \max_{(\tilde{w}_1, \tilde{w}_2)}~\min_{(m,n) \in \mathbb F_{(M_1-1, M_2-1)}^2} d(m,n) \\
&{\rm s.t.~}\frac{b_k}{a_k}< \frac{|\tilde{h}_2| \tilde{w}_2 }{|\tilde{h}_1| \tilde{w}_1}\le \frac{b_{k+1}}{a_{k+1}},
0< \tilde{w}_1 \le 1 {\rm~and~} 0 < \tilde{w}_2 \le 1.
\end{align}
\end{subequations}

By applying the propositions in last subsections, we obtained the following lemma related to the optimal solution to problem~\eqref{eqn:maxminbyinterval}.
\begin{lemma}\label{thm:subinterval}
The optimal solution to~\eqref{eqn:maxminbyinterval} is given as follows:
 \begin{align*}
g\Big(\frac{b_k}{a_k}, \frac{b_{k+1}}{a_{k+1}}\Big)=\begin{cases}
            \frac{|\tilde{h}_2|}{b_k+b_{k+1}}, {\rm~with~} (\tilde{w}_1^*(k), \tilde{w}_2^*(k))=\big(\frac{|\tilde{h}_2| (a_k   +a_{k+1}) }{|\tilde{h}_1| (b_k+b_{k+1})},1\big),  &{\rm if~}\frac{|\tilde{h}_2|}{|\tilde{h}_1|} \le \frac{b_k+b_{k+1}}{a_k+a_{k+1}};\\
            \frac{|\tilde{h}_1| }{a_k+a_{k+1}},{\rm~with~} (\tilde{w}_1^*(k), \tilde{w}_2^*(k))=\big(1, \frac{|\tilde{h}_1| (b_k+b_{k+1})}{|\tilde{h}_2|(a_k+a_{k+1})}\big), &{\rm if~}\frac{|\tilde{h}_2|}{|\tilde{h}_1|} > \frac{b_k+b_{k+1}}{ a_k+a_{k+1}}.
            \end{cases}
\end{align*}

	\hfill$\blacksquare$
\end{lemma}

The proof of Lemma~1 can be found in Appendix\ref{appendix:theorem1}.

Now, we are ready to present the closed-form optimal solution to Problem~\ref{pbm:maxminopt} in terms of $(w_1^*, w_2^*)$ {instead of $(\tilde{w}_1^*, \tilde{w}_2^*)$ defined in~\eqref{eqn:changelabel}} for clarity, which maximizes the minimum Euclidean distance of the sum-constellation, denoted by $d_{\rm noma}$, over the entire feasible region.
\begin{theorem}\emph{Closed-form optimal weighting coefficients:}\label{thm:gaussianmacpower}
The optimal solution to Problem~\ref{pbm:maxminopt} in terms of $(w_1^*, w_2^*)$ is given by:
\begin{align}\label{eqn:optimalweightting}
&(w_1^*, w_2^*)=
\begin{cases}
\big(\sqrt{\frac{3 P_2 M_2^2 }{2 (M_2^2-1) }} \frac{|h_2|}{|h_1|}, \sqrt{\frac{3 P_2}{2(M_2^2-1)}}\big),
 &{\rm if~}\frac{|h_2|}{|h_1|} \le\sqrt{\frac{P_1 (M_2^2-1)}{P_2 M_2^2(M_1^2-1)}};\\
\big(\sqrt{\frac{3 P_1}{2(M_1^2-1)}}, \sqrt{\frac{3 P_1}{2 M_2^2 (M_1^2-1)}} \frac{|h_1|}{|h_2|}\big),
&{\rm if~}\sqrt{\frac{P_1 (M_2^2-1)}{P_2 M_2^2 (M_1^2-1)}}<\frac{|h_2|}{|h_1|} \le \sqrt{\frac{P_1 M_1^2 (M_2^2-1)}{P_2 M_2^2 (M_1^2-1)}};\\
\big(\sqrt{\frac{3 P_2}{2 M_1^2 (M_2^2-1)}} \frac{|h_2|}{|h_1|}, \sqrt{\frac{3 P_2}{2(M_2^2-1)}} \big),
&{\rm if~}\sqrt{\frac{P_1 M_1^ 2 (M_2^2-1)}{P_2 M_2^2 (M_1^2-1)}}<\frac{|h_2|}{|h_1|} \le \sqrt\frac{P_1 M_1^2 (M_2^2-1)}{P_2 (M_1^2-1)};\\
\big(\sqrt{\frac{3 P_1}{2(M_1^2-1)}}, \sqrt{\frac{3 P_1 M_1^2 }{2(M_1^2-1)}} \frac{|h_1|}{|h_2|}\big),
&{\rm if~}\sqrt{\frac{P_1 M_1^2 (M_2^2-1)}{P_2 (M_1^2-1)} }<\frac{|h_2|}{|h_1|}.
\end{cases}
\end{align}

The resulting minimum Euclidean distance $d_{\rm noma}$ in each case is:
\begin{align}\label{eqn:minDist}
&d_{\rm noma}=
\begin{cases}
\sqrt{\frac{3 P_2}{2(M_2^2-1)}}|h_2|, &{\rm if~}\frac{|h_2|}{|h_1|} \le\sqrt{\frac{P_1 (M_2^2-1)}{P_2 M_2^2(M_1^2-1)}};\\
\sqrt{\frac{3 P_1}{2M_2^2 (M_1^2-1)}}|h_1|, &{\rm if~}\sqrt{\frac{P_1 (M_2^2-1)}{P_2 M_2^2 (M_1^2-1)}}<\frac{|h_2|}{|h_1|} \le \sqrt{\frac{P_1 M_1^2 (M_2^2-1)}{P_2 M_2^2 (M_1^2-1)}};\\
\sqrt{\frac{3 P_2}{2 M_1^2 (M_2^2-1)}} |h_2|, &{\rm if~}\sqrt{\frac{P_1 M_1^ 2 (M_2^2-1)}{P_2 M_2^2 (M_1^2-1)}}<\frac{|h_2|}{|h_1|} \le \sqrt\frac{P_1 M_1^2 (M_2^2-1)}{P_2 (M_1^2-1)};\\
\sqrt{\frac{3 P_1}{2(M_1^2-1)}}|h_1|, &{\rm if~}\sqrt{\frac{P_1 M_1^2 (M_2^2-1)}{P_2 (M_1^2-1)} }<\frac{|h_2|}{|h_1|}.
\end{cases}
\end{align}
~\hfill\QED
\end{theorem}

The proof is provided in Appendix\ref{appendix:theoremac}.

\begin{remark}
	 By combing~Eqs.~\eqref{eqn:changelabel} and~\eqref{eqn:optimalweightting}, we can observe that at least one transmitter should transmit with the maximum power. The principle behind this is that we could always scale up both users' transmit powers without violating the power constraint such that the minimum Euclidean distance is enlarged.~\hfill\QED
\end{remark}

We have the following remark regarding the choice of constellation size $M_1, M_2$.
\begin{remark} In order to attain the results in Theorem\,\ref{thm:gaussianmacpower} with the aid of Farey sequence, we assume that $\min\,\{M_1, M_2\} \ge 2$. However, it can be verified that for $M_1=1, M_2\ge 2$ or $M_1\ge 2, M_2=1$, although~\eqref{eqn:optimalweightting} is no longer true,~\eqref{eqn:minDist} still holds.
In fact, if $M_1=1, M_2\ge 2$, we have $(w_1^*, w_2^*)=(0, \sqrt{\frac{3 P_2}{2(M_2^2-1)}})$. Else if $M_1\ge 2, M_2=1$, we have $(w_1^*, w_2^*)=(\sqrt{\frac{3 P_1}{2(M_1^2-1)}},0)$. That is, by assuming $M_k=1$, $k=1,2$, i.e., no information is transmitted by user $S_k$, we should let it keep silent, and thus all the channel resources are allocated to the other user exclusively, who should transmit at its maximum allowable power.
~\hfill\QED
\end{remark}	

We also have the following corollary about the optimal solution described in Theorem~\ref{thm:gaussianmacpower}:
\begin{corollary}\label{cor:uniformconst}
The sum-constellation at the receiver is a \emph{standard} $M_1^2 M_2^2$-QAM constellation with the minimum Euclidean distance $d_{\rm noma}$ affected by the instantaneous channel realizations as given in~\eqref{eqn:minDist}. ~\hfill\QED
\end{corollary}

The proof is provided in Appendix\ref{appendix:cormindist}.

Due to this nice structure of the sum-constellation, the ML decoder reduces to a simple quantizer for the complex constellation \cite{dong16jstsp}, where the detection can be performed for the in-phase and quadrature components separately since they are separable. It is worth mentioning that if $\frac{|h_2|}{|h_1|} \le \sqrt{\frac{P_1 M_1^2 (M_2^2-1)}{P_2 M_2^2 (M_1^2-1)}}$, we have $\frac{|h_1| w_1^*}{|h_2| w_2^*}=M_2$, i.e., the constellation of $S_2$ will have a smaller Euclidean distance than that of $S_1$ at the receiver side; Otherwise if $\frac{|h_2|}{|h_1|} >\sqrt{\frac{P_1 M_1^ 2 (M_2^2-1)}{P_2 M_2^2 (M_1^2-1)}}$, we attain $\frac{|h_1| w_1^*}{|h_2| w_2^*}=\frac{1}{M_1}$, i.e., the constellation of $S_1$ will have a smaller Euclidean distance than that of $S_2$. 

\subsection{The Superiority of NOMA over TDMA}
It is significant to conduct comparisons between NOMA and OMA, such as in~\cite{Poor2017jsac}. Now, to facilitate this comparison  of NOMA over OMA with finite-alphabet inputs, we compare the minimum Euclidean distance of the proposed NOMA and that of TDMA under the same channel realization. In general, for TDMA, the overall available frame is partitioned \emph{uniformly} into orthogonal time slots of the same length for the ease of symbol synchronization. 
Specifically, for a two-user TDMA, we assume that each user has half of the total available time slots and therefore, they should employ $M_1^2$- and $M_2^2$-ary PAM constellations, respectively, to maintain the same transmission rate. In this comparison, we also assume that the channel state of both users remains unchanged {(i.e., quasi-static)} during the two consecutive time slots.

For TDMA, the minimum Euclidean distance for users $S_1$ and $S_2$ are $d_{\rm oma,1}=\sqrt{\frac{3 P_1}{2(M_1^4-1)}}|h_1|$ and $d_{\rm oma,2}=\sqrt{\frac{3 P_2}{2(M_2^4-1)}}|h_2|$, respectively. Now, we denote the minimum Euclidean distance among the two users as:
\begin{align}\label{eqn:minDistoma}
	d_{\rm oma}=\min\,\{d_{\rm oma,1}, d_{\rm oma,2}\}=\min\,\Big\{\sqrt{\frac{3 P_1}{2(M_1^4-1)}}|h_1|, \sqrt{\frac{3 P_2}{2(M_2^4-1)}}|h_2| \Big\}.
\end{align}

We then have the following corollary regarding the resulting minimum Euclidean distance of both schemes:
\begin{corollary}\label{cor:distnomavsoma}
The minimum Euclidean distance of the proposed NOMA, $d_{\rm noma}$ given in~\eqref{eqn:minDist}, is strictly larger than that of the TDMA scheme, $d_{\rm oma}$ given in~\eqref{eqn:minDistoma}, with equal time-slot allocation. That is, $d_{\rm noma}>d_{\rm oma}$ holds for arbitrary given channel realizations $h_1$, $h_2$ and constellation sizes $M_1$, $M_2$.\hfill\QED
\end{corollary}	

The proof is provided in Appendix\ref{appendix:compnomaoma}. From Corollary~\ref{cor:distnomavsoma}, since $d_{\rm noma}>d_{\rm oma}$, it is expected that NOMA outperforms TDMA in terms of error performance, especially in moderate and high SNR regions as can be confirmed by numerical results.

\section{Rate Allocation in Two-User Gaussian MAC with Sum-Rate Constraint}\label{sec:optimal_rate_allocation}
In this section, we consider the optimal rate-allocation problem among the two users under a sum-rate constraint for the  above two-user Gaussian MAC with a finite PAM constellation. Moreover, a high-rate asymptotically optimal solution is also provided when the transmission rates of both users are relatively high.
\subsection{Problem Formulation}
{From a radio resource management perspective, when user fairness is not a major concern, one of the most important problems is to maximize the minimum Euclidean distance at the receiver side, which determines the system error performance in moderate and high SNR regimes.
This motivates us to consider the maximization of $d_{\rm noma}$ in~\eqref{eqn:minDist} under a sum-rate constraint.}
Mathematically speaking, we intend to solve the following optimization problem:

\begin{problem}\emph{Rate allocation in two-user MAC under a sum-rate constraint:}\label{pbm:rateallo} We aim to maximize the minimum Euclidean distance of the received sum-constellation $d_{\rm noma}$ given in~\eqref{eqn:minDist} by adjusting the constellation sizes of both users under a sum-rate constraint, that is:
\begin{align}\label{eqn:rateallo}
	\max_{M_1, M_2}~d_{\rm noma} \quad {\rm~s.t.~} \log_2 M_1+\log_2 M_2 =\log_2 M,
\end{align}
where $M$ is the size of the sum-constellation, and $\log_2 M_1, \log_2 M_2 \in \mathbb N_{\log_2 M}$ ({i.e., $M_1$, $M_2$ are non-negative integer powers of 2}).
~\hfill\QED
\end{problem}

\subsection{Optimal Rate Allocation}
In this subsection, we investigate the above rate-allocation problem. From~\eqref{eqn:rateallo}, we have $M_2=M/M_1$, and therefore the minimum distance $d_{\rm noma}$ can be considered as a piecewise function of $M_1$ for any given $\lambda=\frac{P_2 |h_2|^2}{P_1 |h_1|^2}$. Note that, in~\eqref{eqn:rateallo}, the channel coefficients $|h_1|$, $|h_2|$ and power constraints $P_1$, $P_2$ are treated as constant, and thus $\lambda$ is also considered as a known constant. Then, Problem~\ref{pbm:rateallo} is equivalent to the minimization of $\beta(M_1)=\frac{3 P_1 |h_1|^2}{2 d_{\rm noma}^2}$ and with the help of~\eqref{eqn:minDist}, we can attain:
\begin{align}\label{eqn:betafunction}
\beta(M_1)=\begin{cases}
\frac{1}{\lambda}\big(\frac{M^2}{M_1^2}-1\big), & {\rm if~} 1\le M_1 \le \gamma_1(\lambda);\\
M^2-\frac{M^2}{M_1^2}, &{\rm if~} \gamma_1(\lambda) < M_1 \le \gamma_2(\lambda);\\
\frac{1}{\lambda}(M^2-M_1^2), &{\rm if~} \gamma_2(\lambda)< M_1 \le \gamma_3(\lambda);\\
M_1^2-1, &{\rm if~} \gamma_3(\lambda)< M_1 \le M,
\end{cases}
\end{align}
where $\gamma_1(\lambda)=\sqrt{\frac{\lambda+1}{\lambda+ \frac{1}{M^2}}}$, $\gamma_2(\lambda)= \sqrt{\frac{\sqrt{(\lambda-1)^2 + \frac{4\lambda}{M^2}} -(\lambda-1)}{2}} M$, and $\gamma_3(\lambda)=\sqrt{\frac{\lambda+M^2}{\lambda+1}}$.

From~\eqref{eqn:betafunction}, we can find that, $\beta(M_1)$ is  monotonically decreasing for $1\le M_1 \le \gamma_1(\lambda)$ and $\gamma_2(\lambda)< M_1 \le \gamma_3(\lambda)$, and it is monotonically increasing for $\gamma_1(\lambda) < M_1 \le \gamma_2(\lambda)$ and $\gamma_3(\lambda)< M_1 \le M$. Consequently, the optimal rate-allocation solution can be stated as follows:
\begin{theorem}
The optimal solution to Problem\,\ref{pbm:rateallo} is $M_1^*=M_{1,(k^*)}$ and $M_2^*=M/M_1^*$ such that $k^*=\arg \min \beta_k $, in which $\beta_1=\frac{1}{\lambda}\big(\frac{M^2}{M_{1,(1)}^2}-1\big)$, where $M_{1,(1)}=2^{\lfloor \log_2 \gamma_1(\lambda) \rfloor }$;
$$\beta_2=\begin{cases}M^2-\frac{M^2}{M_{1,(2)}^2 }, &{\rm~if~} \lfloor \log_2 \gamma_1(\lambda) \rfloor \le \lfloor \log_2 \gamma_2(\lambda) \rfloor +1\\
     \infty, &{\rm otherwise}\end{cases}, {\rm~where~}M_{1,(2)}=2^{\lfloor \log_2 \gamma_1(\lambda) \rfloor +1};$$
$$\beta_3=\begin{cases}\frac{1}{\lambda}(M^2-M_{1,(3)}^2), &{\rm~if~} \lfloor \log_2 \gamma_2(\lambda) \rfloor \le \lfloor \log_2 \gamma_3(\lambda) \rfloor +1\\
\infty, &{\rm otherwise}\end{cases}, {\rm~where~} M_{1,(3)}=2^{ \lfloor \log_2 \gamma_3(\lambda) \rfloor};$$
and $\beta_4= M_{1,(4)}^2-1$, where $M_{1,(4)}=2^{\lfloor \log_2 \gamma_3(\lambda) \rfloor +1}$, {where $\lfloor a\rfloor$ is the floor function which returns the largest integer no more than $a$.}	
\hfill$\Box$
\end{theorem}

Although the above solution is optimal, the structure of the sum-constellation as a function of $\lambda$ is not straightforward enough. In the following part, we will give an asymptotically optimal solution to draw some useful insights and also reveal the merits of the proposed NOMA scheme.
\subsection{Asymptotically Optimal Solution when the Transmission Rates of Both Users Are High}\label{subsec:asympt_rate_allocation}
For the considered two-user system, the most interesting case is when the transmission rates of both users go to infinity~\cite{Tse08,Tse11April}. In a such case, we have $\lim_{M_1\to\infty} \frac{M_1^2-1}{M_1^2} =1$ and $\lim_{M_2\to\infty} \frac{M_2^2-1}{M_2^2}=1$ and then
 $\beta (M_1)$ in\,\eqref{eqn:betafunction} will converge to $\tilde \beta(\tilde M_1)$, such that:
\begin{align}\label{eqn:approximatemindist}
\tilde{\beta}(\tilde{M}_1)=
\begin{cases}
\frac{M^2}{\lambda \tilde{M}_1^2}, &{\rm if~} 1\le \tilde{M}_1 \le \frac{1}{\sqrt{\lambda}};\\
M^2, &{\rm if~} 1 \le \frac{1}{\sqrt{\lambda}} < \tilde{M}_1 \le M;\\
\frac{M^2}{\lambda}, &{\rm if~} \frac{1}{M}\le \frac{\tilde{M}_1}{M} \le \frac{1}{\sqrt{\lambda}}<1;\\
\tilde{M}_1^2, &{\rm if~} \frac{1}{\sqrt{\lambda}}<\frac{\tilde{M}_1}{M} \le 1.
\end{cases}
\end{align}

\begin{problem}[Asymptotically optimal rate-allocation problem in MAC under a sum-rate constraint]\label{pbm:asympopt}
We intend to solve the following optimization problem by adjusting $\tilde{M}_1$ and $\tilde{M}_2$ subject to a sum-rate constraint, given by:
	\begin{align}
	\min_{\tilde{M}_1, \tilde{M}_2}~\tilde{\beta}(\tilde{M}_1)\quad {\rm~s.t.~} \log_2 \tilde{M}_1+\log_2 \tilde{M}_2 =\log_2 M,
	\end{align}
	where $\tilde{\beta}(\tilde{M}_1)$  is defined in~\eqref{eqn:approximatemindist},  and $\tilde{M}_1$, $\tilde{M}_2$ are powers of 2.
	~\hfill\QED
\end{problem}

Since the objective function is a simple piecewise function of $\tilde{M}_1$, we are ready to formally give our solutions with no need of proof:
\begin{theorem}\emph{Asymptotically optimal solution:}\label{thm:asymprateallo}
The asymptotically optimal solution to Problem~\ref{pbm:asympopt} is given by:
\begin{enumerate}
	\item If $\lambda \le 1$, we have $\tilde{M}_1^*=\min\,\{2^{\lfloor \log_2 \frac{1}{\sqrt{\lambda}} \rfloor +1 }, M \}$
	  and  $\tilde{M}_2^*=\frac{M}{\tilde{M}_1^*}$;
	\item If $\lambda >1$, we have $\tilde{M}_1^*=\max\,\{2^{\lfloor \log_2\frac{M}{\sqrt{\lambda}} \rfloor}, 1\}$
	and $M_2^*=\frac{M}{\tilde{M}_1^*}$.~\hfill \QED
\end{enumerate}
\end{theorem}

{
We have the following remark on the asymptotically optimal solution:
\begin{remark}
We consider the case in a high SNR regime and with near-far effect such that $\frac{1}{2}\log (1+\frac{P_1 |h_1|^2}{2\sigma^2}) \gg \frac{1}{2}\log (1+\frac{P_2 |h_2|^2}{2\sigma^2}) \gg 1$. We then have the following two cases:
\begin{itemize}
 \item {\bf Case 1}: The sum-rate is relatively low such that $M^2\le\frac{1}{\lambda}$:
 In this case, with Theorem~\ref{thm:asymprateallo}, we have  $\tilde M_1^* = \min\,\{2^{\lfloor \log_2 \frac{1}{\sqrt{\lambda}} \rfloor +1 }, M \}= M$ and $ \tilde{M}_2^*=M/\tilde M_1^*=1$. In other words, the channel should be solely allocated to the user with stronger channel, and our scheme degrades into the OMA method with the resulting minimum Euclidean distance $d_{\rm noma}= \sqrt{\frac{3 P_1}{2 M^2}}|h_1|$.
 \item {\bf Case 2}: The sum-rate is high enough such that $M^2>\frac{1}{\lambda}$:
 Likewise, by Theorem~\ref{thm:asymprateallo}, we have $\tilde M_1^* = \min\,\{2^{\lfloor \log_2 \frac{1}{\sqrt{\lambda}} \rfloor +1 }, M \} =  2^{\lfloor \log_2 \frac{1}{\sqrt{\lambda}} \rfloor +1 } \approx  \frac{1}{\sqrt{\lambda}} =\frac{\sqrt{P_1} |h_1|}{\sqrt{P_2} |h_2|}$ and
 $\tilde{M}_2^*=M/\tilde{M}_1^* \approx \frac{\sqrt{P_2} |h_2|}{\sqrt{P_1} |h_1|} M$.
 Now, by~\eqref{eqn:minDist}, we can attain $d_{\rm noma}=\sqrt{\frac{3 P_1}{2 \tilde{M}_2^{*2} (\tilde{M}_1^{*2}-1)}}|h_1|
 {\approx}\sqrt{\frac{3 P_1}{2 M^2}}|h_1|$. 
\end{itemize}
Overall, for the proposed NOMA design in a high SNR regime and with near-far effect, if the sum-rate is low, the weak user should keep silent and the channel resources are solely allocated to the user with a stronger channel. On the other hand, if the sum-rate is high enough, the minimum Euclidean distance of the proposed NOMA design is close to that when only the strong user transmits free of interference. This means that in our scheme, the support of the weak user to transmit at a non-zero rate together with the strong user will cause almost no degradation to the system performance. But, the rate of the weak user highly depends on the near-far effect.
\end{remark}}


\section{Simulation Results and Discussions}
In this section, we carry out computer simulations to verify the effectiveness of our NOMA design relative to the CR-NOMA design proposed in~\cite{Harshan11} and the OMA methods including TDMA and frequency-division multiple access (FDMA) schemes in various channel conditions and system configurations. More specifically, we consider both cases where the transmission rates are fixed or adaptive to channel states. Without loss of generality, we assume that $P_1=P_2=1$ and the system signal-to-noise ratio (SNR) is defined by $\rho \triangleq 1 / {2 \sigma^2}$.  All channels are subject to Rayleigh fading such that $h_k \sim \mathcal{CN}(0, 2\delta_k^2)$, $k=1,2$.
\subsection{Comparison of Average Error Performance with Fixed Transmission Rate}
\begin{figure}
	\centering
	\subfigure[]{
		\includegraphics[width=0.45\linewidth]{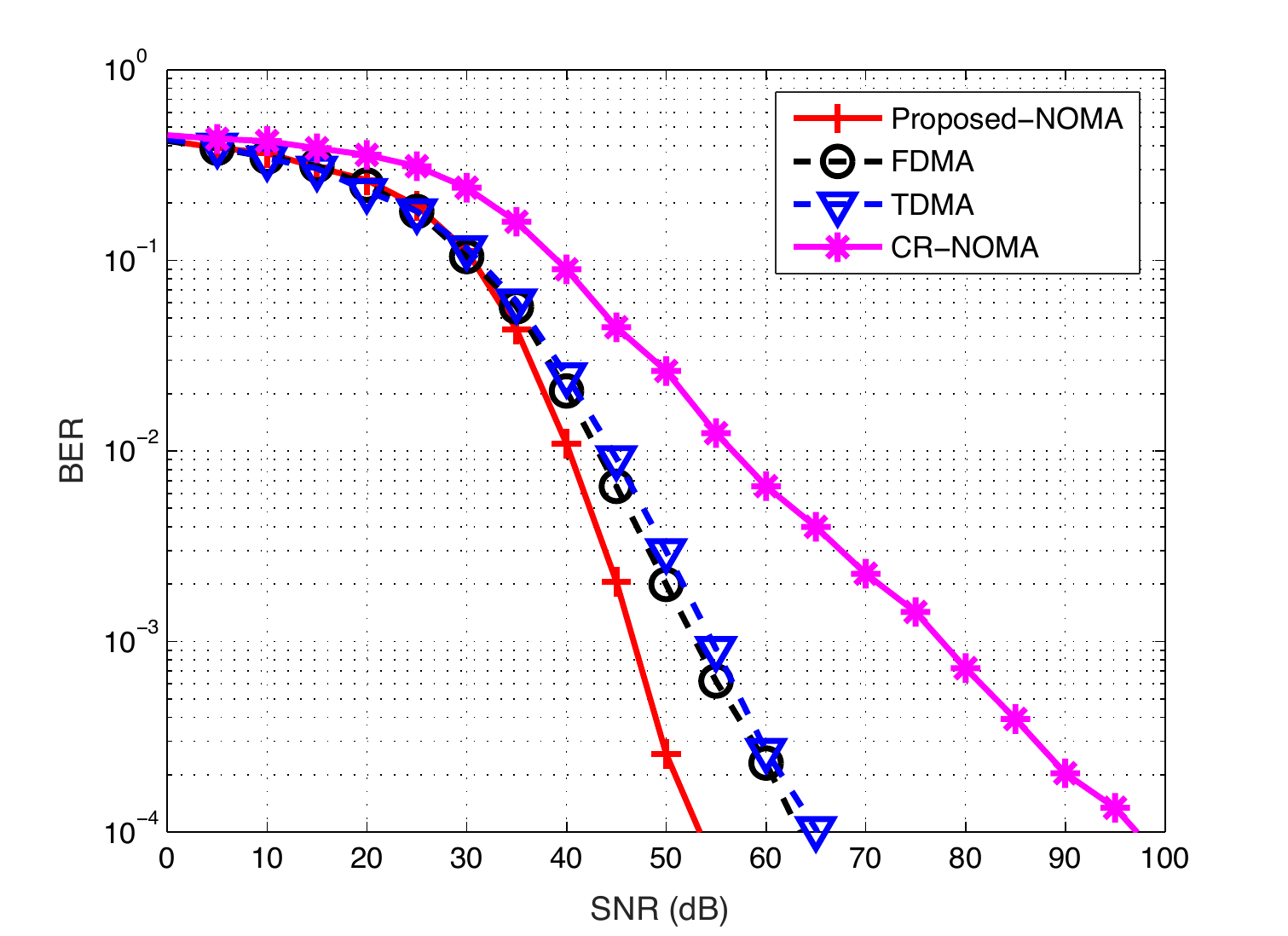}
		\label{fig:FixedRat_equalgain}
	}
	\subfigure[]{
		\includegraphics[width=0.45\linewidth]{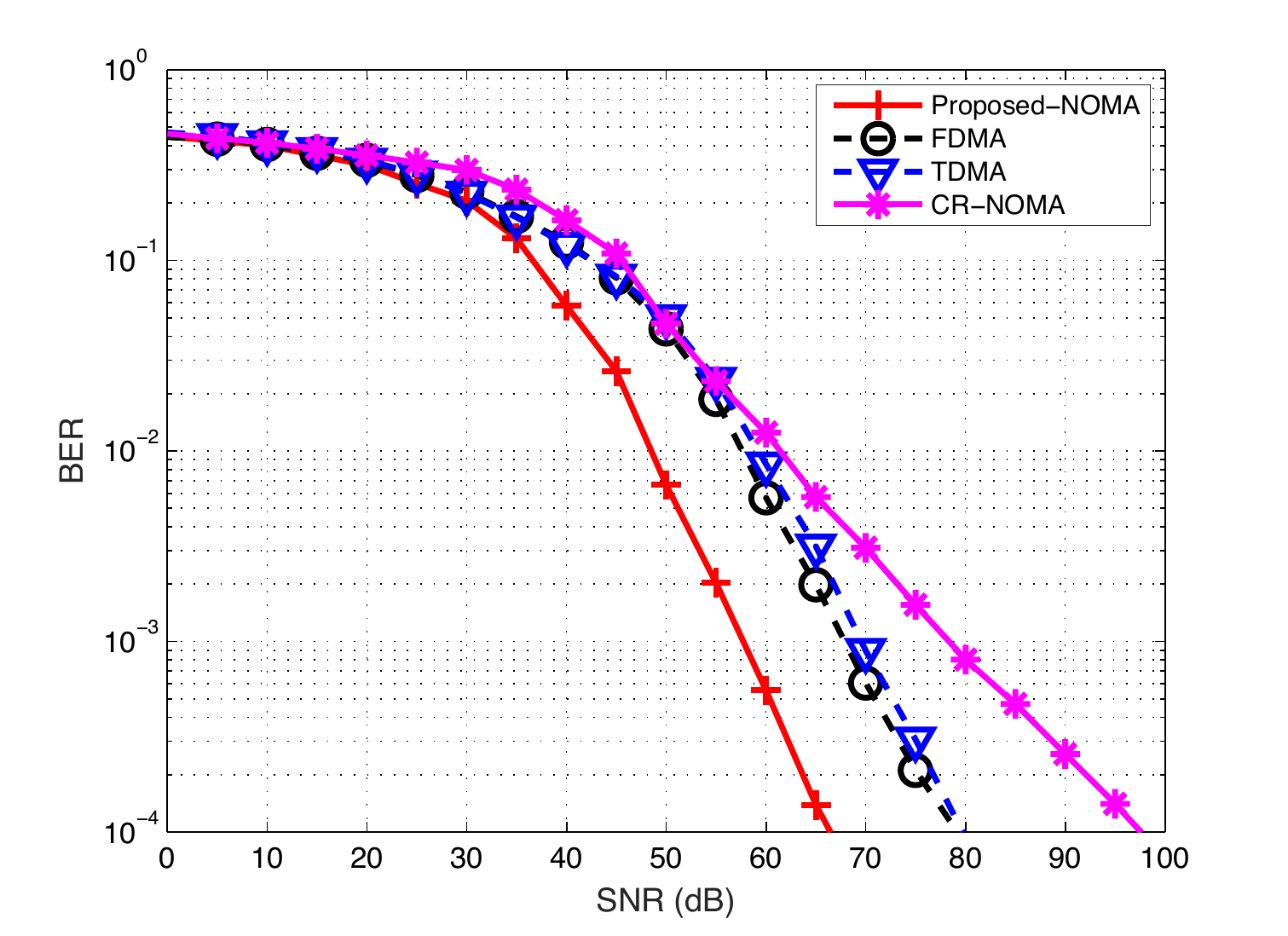}
		\label{fig:FixedRat_unequalgain}
	}
	\vspace{-10pt} \caption{Comparison between the Proposed-NOMA, CR-NOMA, TDMA and FDMA methods where 64-QAM is used for our case and 64-PSK is used for CR-based method: (a) $(\delta_1^2, \delta_2^2)=(1,1)$,~(b) $(\delta_1^2, \delta_2^2)=(1,1/64)$.}
	\label{fig:fixed rate}
\end{figure}
We first consider the case where the transmission rates of both users are predetermined, i.e., the constellation sizes $M_1^2$ and $M_2^2$ are fixed. For convenience, we assume that both users transmit alternatively by using half of the total time slots or half of the available frequency band in TDMA and FDMA, respectively. In both methods, to maintain the same data rate for each user, we should increase the constellation sizes by using $M_1^4$- and $M_2^4$-ary QAM constellations instead.  There is no interference occurring at the receiver side since the channels are orthogonal to each other. It is worth mentioning that, for both TDMA and FDMA, we assume that the instantaneous transmit power of each user remains the same as in the NOMA. For FDMA method, as the total bandwidth of each user is halved, the equivalent noise at the receiver side also reduces by half compared with the TDMA method. {Therefore, we would expect the FDMA scheme has an around 3dB SNR gain compared with TDMA method.} For the  CR-NOMA, we let each user transmit at the maximum allowable power by using constellations $\big\{\exp(\frac{j 2\pi k}{N})\big\}_{k=0}^{N-1}$ and $\big\{\exp(\frac{j 2\pi k +j\pi}{N})\big\}_{k=0}^{N-1}$ as proposed in~\cite{Harshan11} for users $S_1$ and $S_2$, respectively.

We first compare the average BER of all the schemes where the variances of the channel coefficients are the same, i.e., $(\delta_1^2, \delta_2^2)=(1,1)$ in Fig.\,\ref{fig:FixedRat_equalgain}. In the simulation,  without loss of generality, we assume that each user adopts $64$-QAM for the proposed NOMA  design and 64-PSK is used by each user in CR-NOMA. Meanwhile, for TDMA and FDMA methods, each user uses 4096-QAM.  As can be observed from Fig.\,\ref{fig:FixedRat_equalgain} that, the proposed NOMA design outperforms all the designs in moderate and high SNR regimes. In addition, the FDMA method has a better error performance than the TDMA scheme as expected. The CR-NOMA has the highest BER due to the fact that the PSK constellation has a smaller Euclidean distance under the same power constraint compared with QAM constellation.

In the following simulation, we take the near-far effect into consideration by letting $(\delta_1^2, \delta_2^2)=(1,1/64)$ as shown in Fig.\,\ref{fig:FixedRat_unequalgain}. Likewise, the proposed NOMA design has the lowest BER compared with all the benchmark schemes.
Also, we can observe that the gap between the proposed NOMA and the FDMA as well as TDMA is larger than that in the case of equal channel gain. For example, at the BER $10^{-3}$, the proposed NOMA has {around} 5dB SNR gain in Fig.\,\ref{fig:FixedRat_equalgain}, while the SNR gain is approximately 10dB in Fig.\,\ref{fig:FixedRat_unequalgain}. Interestingly, we also observe that the error performance of CR-NOMA improves substantially compared to TDMA and FDMA in this case with near-far effect.

From both Figs.\,\ref{fig:FixedRat_equalgain} and~\ref{fig:FixedRat_unequalgain}, we can observe that the performance gain of NOMA is highly related to the relative strength of the channel coefficients. To show this phenomenon clearly, we now study the BER against the relative strength of the channel coefficients under different SNRs. More specifically,  in Fig.\ref{fig:FixedRat_vDelta2SNR40}, we set the variance of user $S_1$ as $\delta_1^2=1$, and we plot the BER against the variance of user $S_2$, i.e., $\delta_2^2$, in dB. It can be observed from Fig.\,\ref{fig:FixedRat_vDelta2SNR40} that, for $\rho=40$dB (i.e., the SNR is relatively low relative to the target transmission rate), our proposed NOMA scheme outperforms all the benchmark schemes. When $\delta_2^2$ is less than 1 (i.e., less than 0dB), the error performance is mainly limited by user $S_1$ and even if $\delta_2^2$ equals to 1, the BER gain of the proposed NOMA method is still marginal. However, with the increase of $\delta_2^2$, the BER gain of the proposed NOMA method increases and finally gets saturated. Actually, when $\delta_2^2$ is extremely large, the BER of the proposed NOMA is close to the system with one user transmitting with 64-QAM in both orthogonal blocks, while for the OMA method, it saturates as one user transmits using 4096-QAM in one block. This validates our observation that the proposed NOMA has a higher SNR gain when there is near-far effect. With the increase of $\delta_2^2$, the performance of CR-NOMA improves dramatically and it eventually outperforms the OMA methods. However, the BER performance is poor when the channel gains of the two users are close. This is {due to the fact that with the same spectral efficiency, a PSK constellation has a smaller minimum Euclidean distance than a QAM constellation. Moreover, the sum-constellation of two PSK constellations at the receiver does not have a good geometric structure,In Fig.\,\ref{fig:FixedRat_vDelta2SNR50}, we can see that with the near-far effect, the BER gain of the proposed NOMA also become more significant. The BER gain of the proposed NOMA is evident even if $\delta_2^2=1$, which coincides well with the {phenomenon observed in Fig.\,\ref{fig:fixed rate}.}

\begin{figure}
	\centering
	\subfigure[]{
		\includegraphics[width=0.48\linewidth]{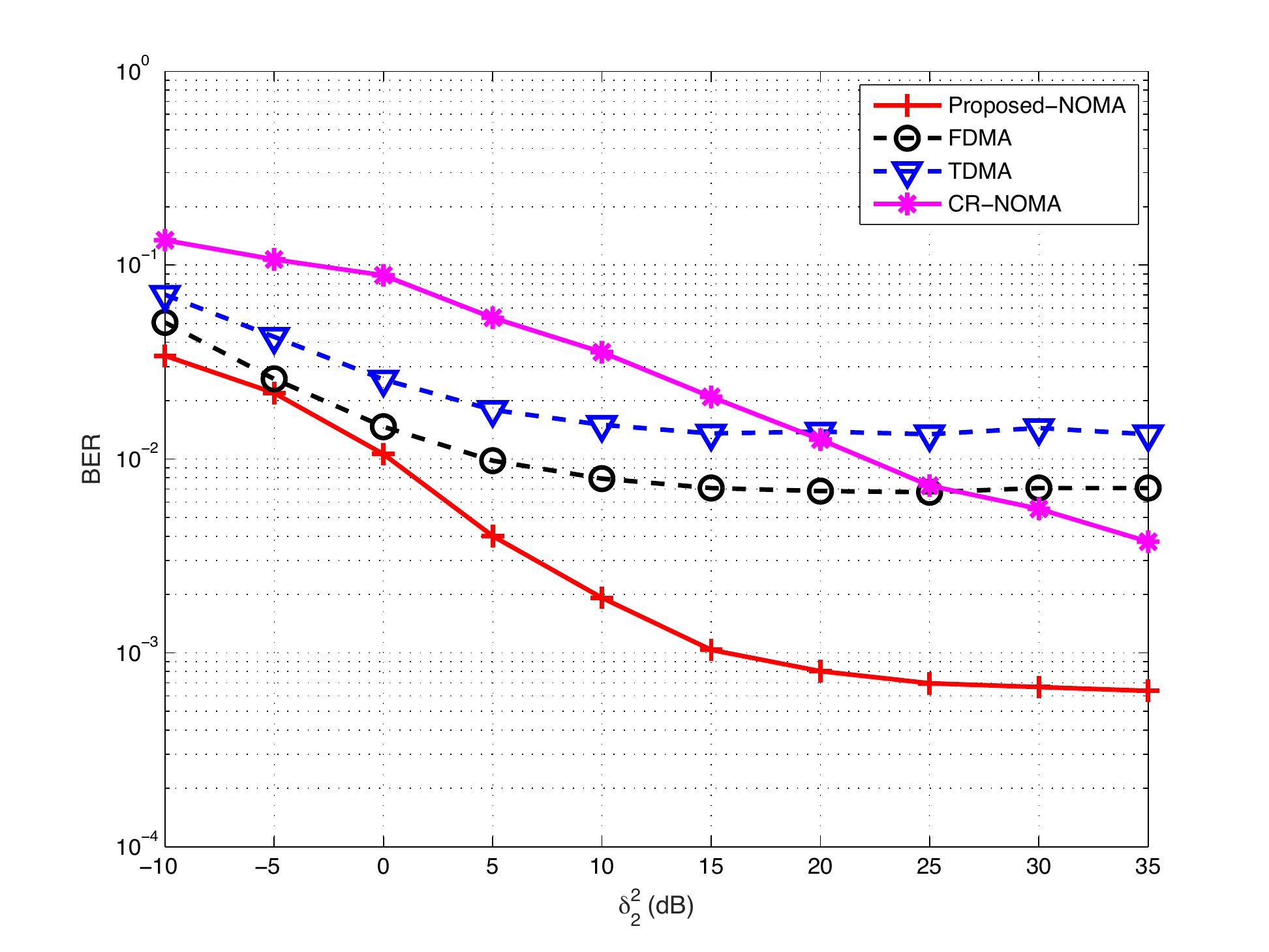}
		\label{fig:FixedRat_vDelta2SNR40}
	}
	\subfigure[]{
		\includegraphics[width=0.45\linewidth]{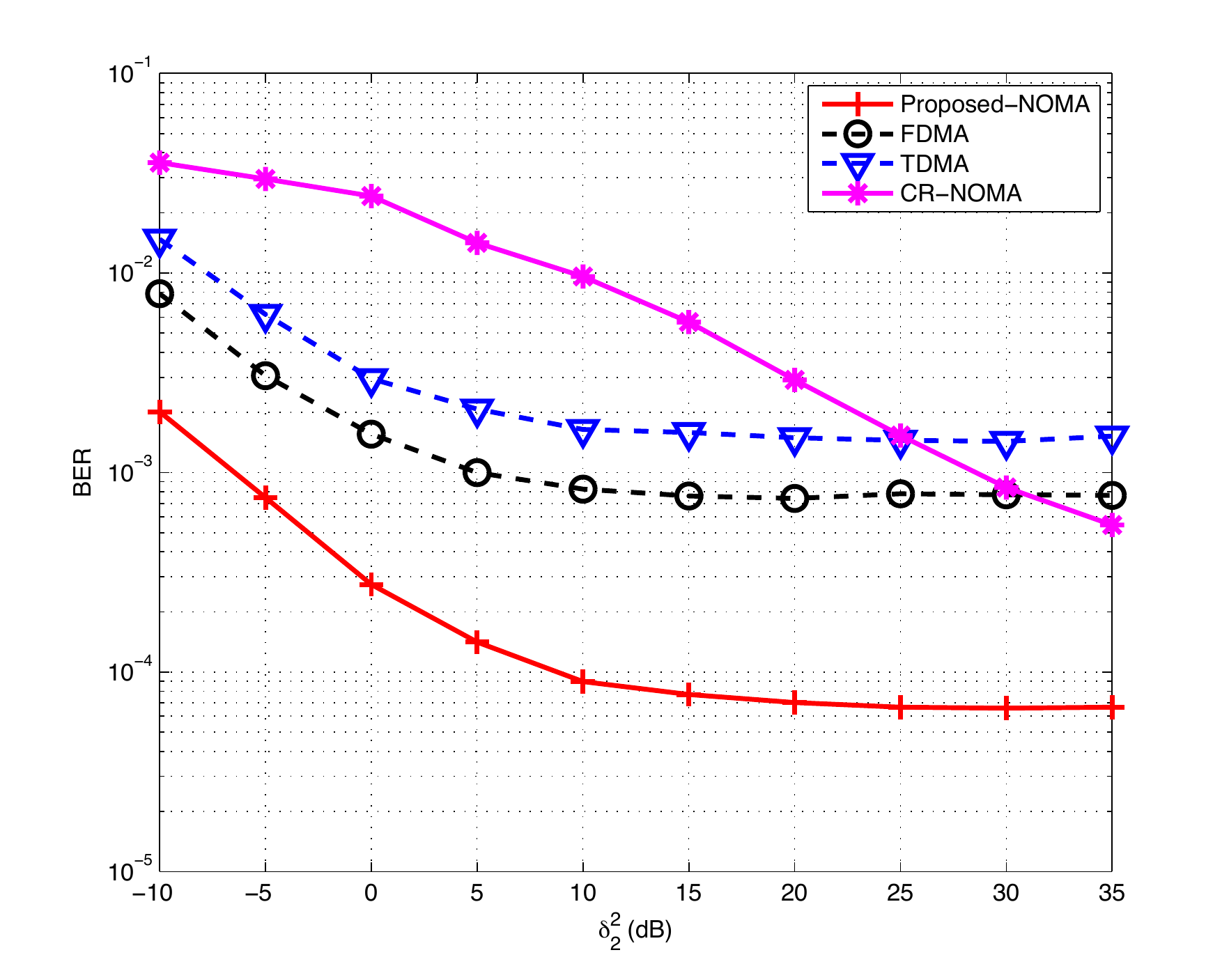}
		\label{fig:FixedRat_vDelta2SNR50}
	}
	\vspace{-10pt} \caption{ Comparison between the Proposed-NOMA with CR-NOMA, TDMA, and FDMA methods, 64-QAM are used for our case and 64-PSK are used for CR based method with (a) $\rho=40$dB.~(b) $\rho=50$dB.}
\end{figure}

\begin{figure}
	\centering
	\subfigure[]{
		\includegraphics[width=0.45\linewidth]{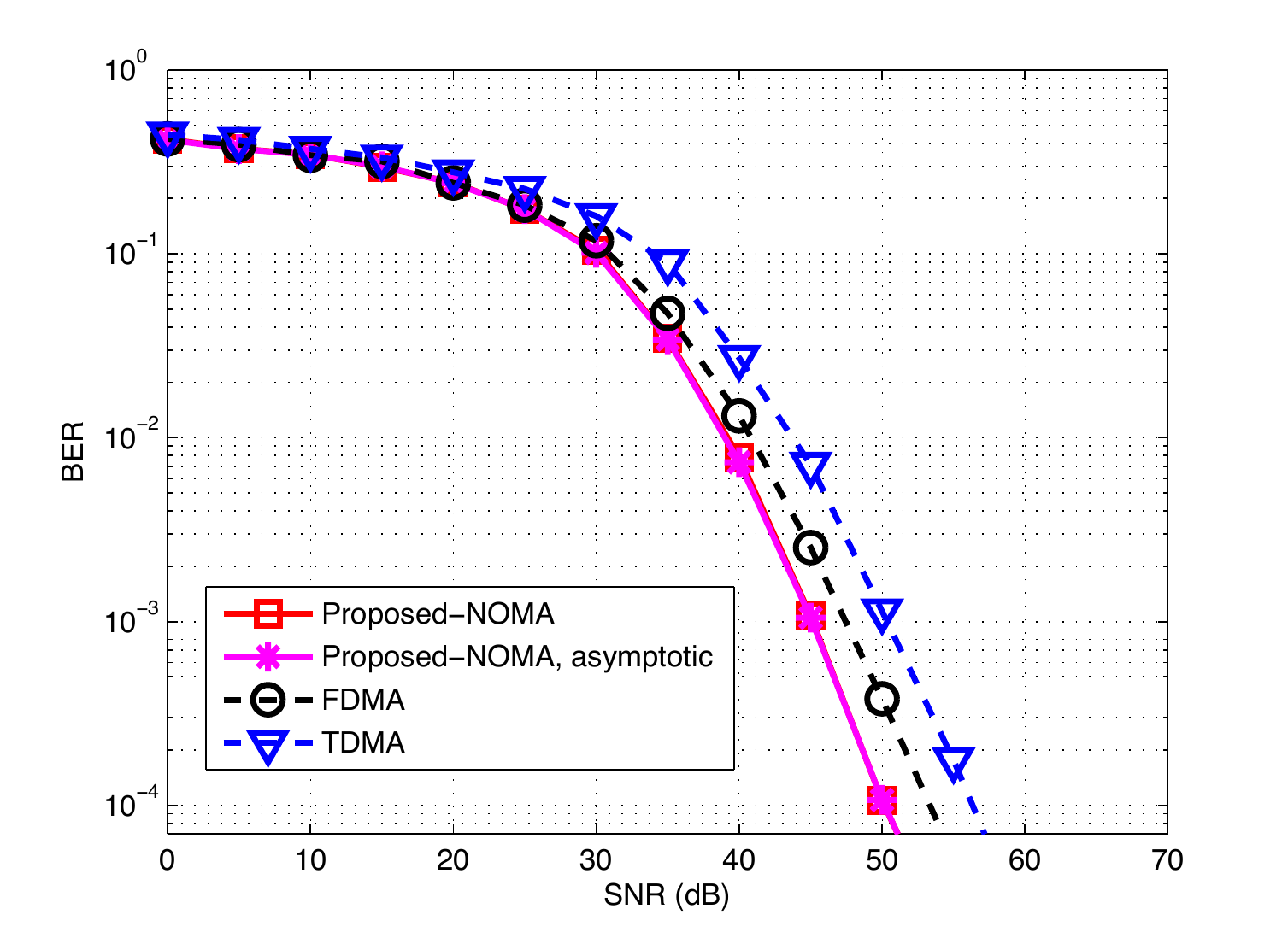}
		\label{fig:adaptiverate_equalgain}
	}
	\subfigure[]{
		\includegraphics[width=0.45\linewidth]{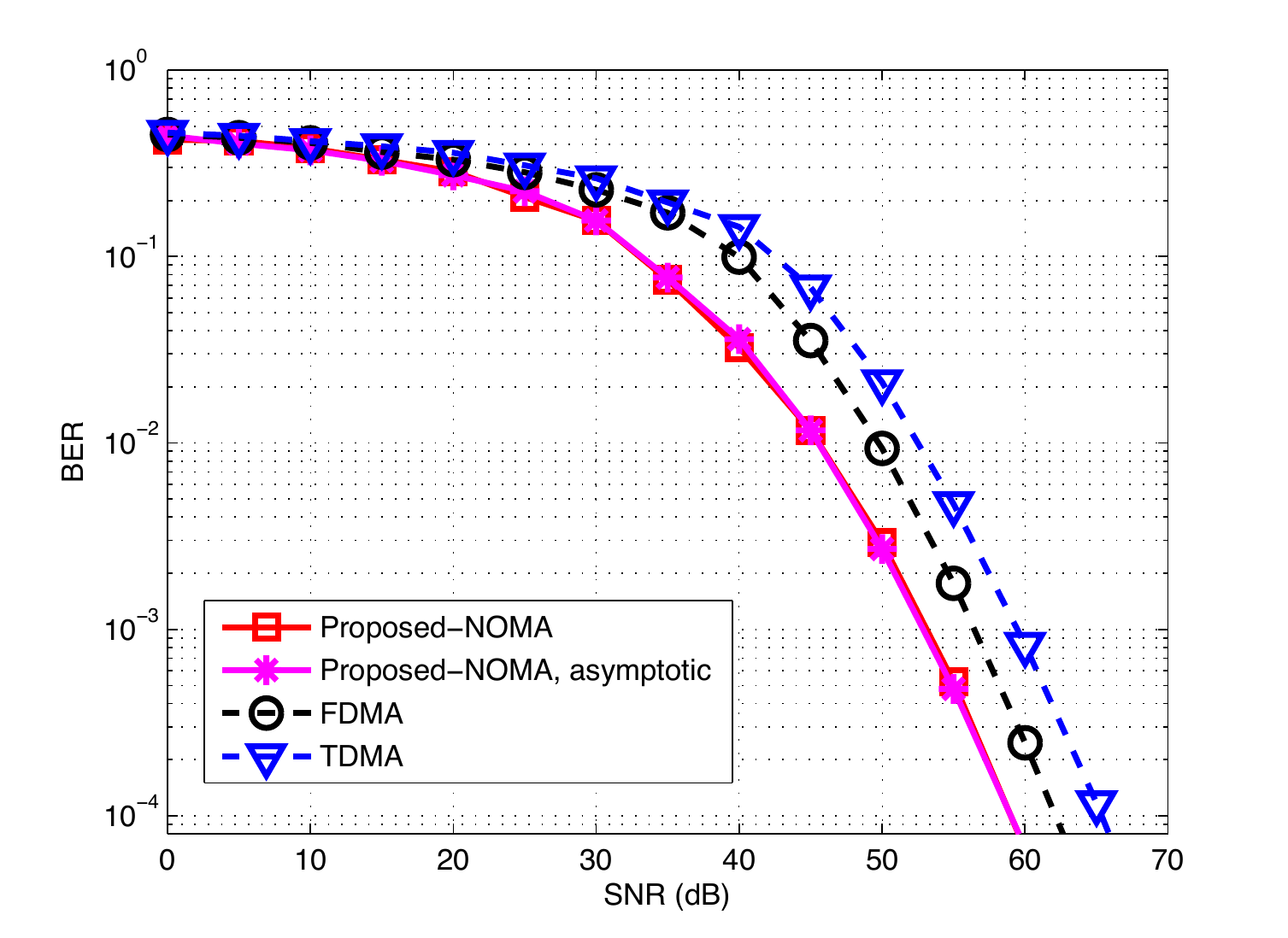}
		\label{fig:adaptiverate_unequalgain}
	}
	\vspace{-10pt} \caption{Comparison between proposed-NOMA, TDD and FDD methods, 64-QAM are used with $M=64$. (a) $(\delta_1^2, \delta_2^2)=(1,1)$, (b) $(\delta_1^2, \delta_2^2)=(1,1/64)$.}
	\label{fig:adaptiverate}
\end{figure}

\begin{figure}
	\centering
	\subfigure[]{
		\includegraphics[width=0.45\linewidth]{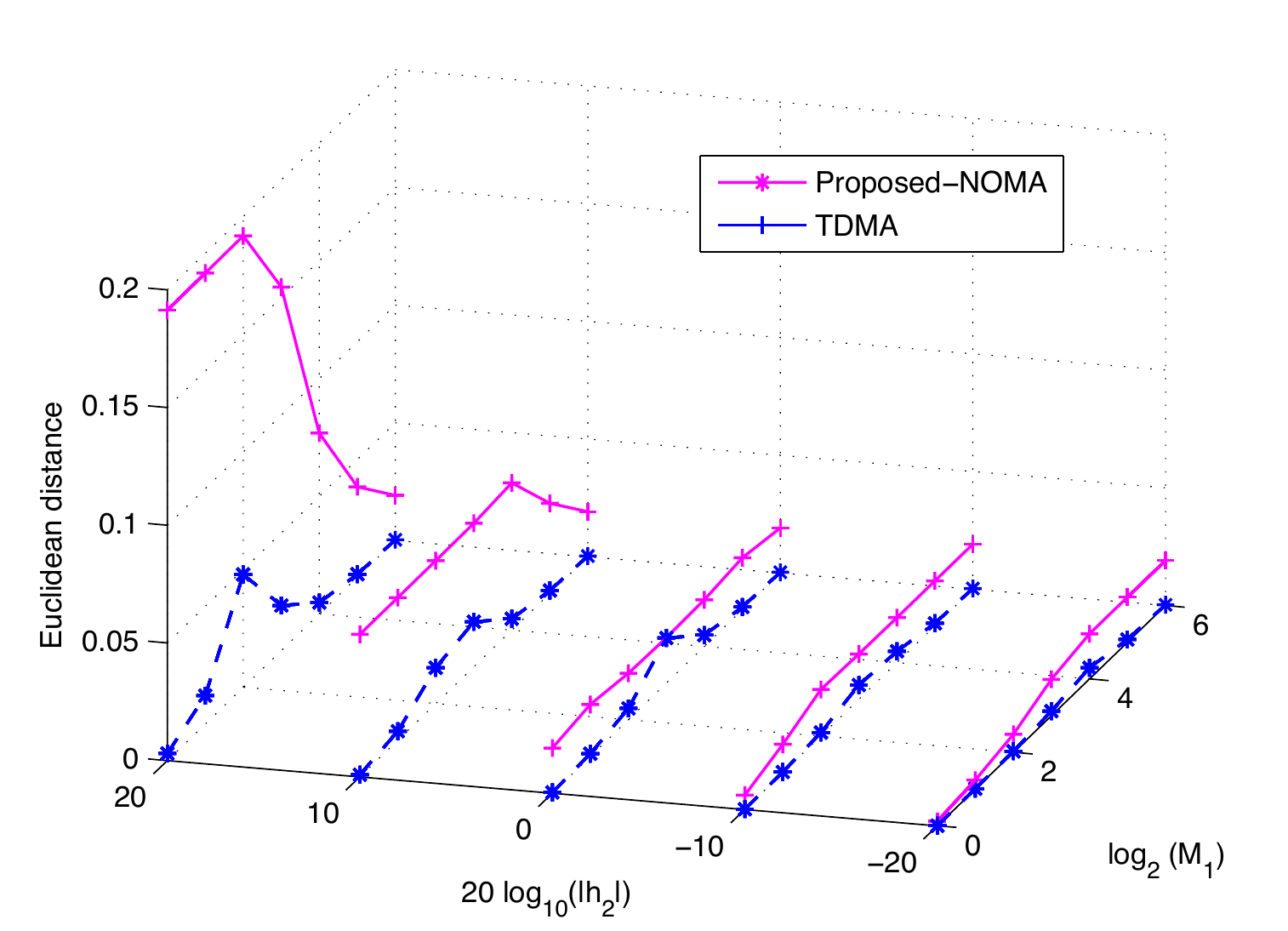}
		\label{fig:Euclidean_NOMAvsOMA_M64}
	}
	\subfigure[]{
		\includegraphics[width=0.45\linewidth]{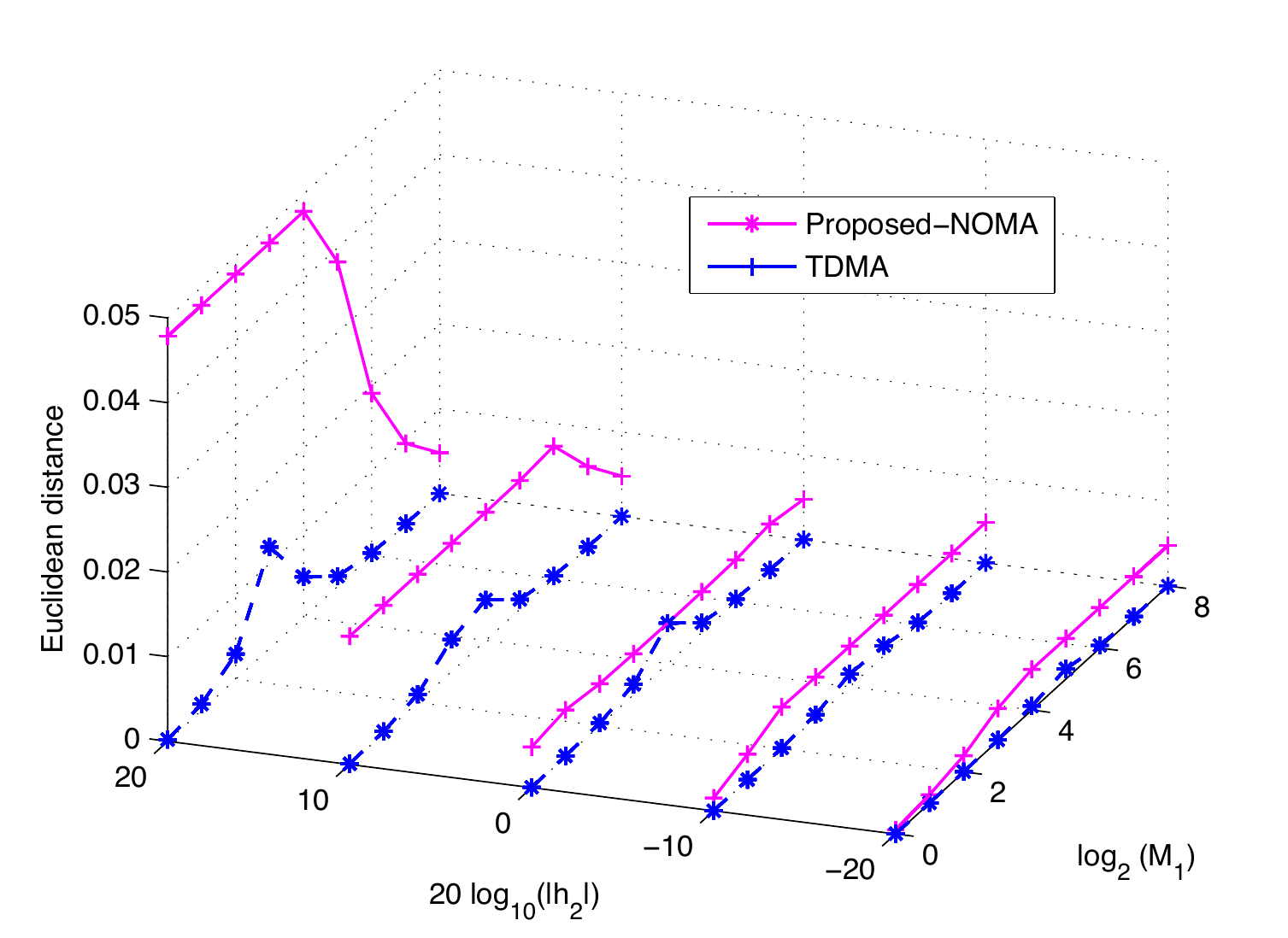}
		\label{fig:Euclidean_NOMAvsOMA_M256}
	}
	\vspace{-10pt} \caption{Comparison of the minimum Euclidean distance of the proposed-NOMA (i.e., $d_{\rm noma}$ given in ~\eqref{eqn:minDist}) and that of TDMA method ($d_{\rm oma}$ given in~\eqref{eqn:minDistoma}) with $|h_1|=1$ for different $|h_2|$ and $M_1$  (a) $M=64$, (b) $M=256$.}
	\label{fig:Euclidean}
\end{figure}

\subsection{Comparison of Average Error Performance with Optimal Rate Allocation}
We now compare the average error performance of our proposed NOMA using the optimal rate allocation (see Sec. \ref{sec:optimal_rate_allocation}) with that of all the benchmark schemes. Note that for TDMA and FDMA, we should find the optimal values of $M_1$ and $M_2$ satisfying $M=M_1 M_2$, to maximize $d_{\rm oma}$ given in~\eqref{eqn:minDistoma}. Without loss of generality, we set $M=64$ and we consider both cases without and with near-far effect as plotted in Figs~\ref{fig:adaptiverate_equalgain} and~\ref{fig:adaptiverate_unequalgain}, respectively. As shown and explained in \cite{Ding16wclrelay}, the superiority of NOMA over OMA can be reduced when both of them adopt the optimal resource allocation. Here, we make similar observations by comparing Fig.~\ref{fig:adaptiverate} with Fig.~\ref{fig:fixed rate}. Nevertheless, as showed in Fig.~\ref{fig:adaptiverate}}, {NOMA yields a considerable performance gain compared to OMA even they both employ the optimal resource allocation, and this gain can further be enlarged by the near-far effect. For example, at the BER of $10^{-3}$, the proposed NOMA has about 3dB SNR gain without near-far effect, which increases to 5dB with near-far effect.} Moreover, we observe that the performance achieved by the asymptotically optimal rate allocation tightly approaches that of the optimal rate allocation solution, which validates our analysis in Sec. \ref{subsec:asympt_rate_allocation}.

\subsection{Comparison of the Minimum Euclidean Distance}
At last, we compare the minimum Euclidean distance of the proposed NOMA design with that of TDMA method in Fig.~\ref{fig:Euclidean}. It can be observed that our proposed NOMA design achieves larger minimum Euclidean distance than TDMA method in all simulated cases, which validates the result presented in Corollary~\ref{cor:distnomavsoma}. We can also see that the stronger the near-far effect is, the larger the performance gap between the proposed NOMA and TDMA. More importantly, for the proposed NOMA, when we enlarge $M_1$, there is a large interval in which the minimum Euclidean distance of NOMA (i.e., $d_{\rm noma}$) will remain almost unchanged, while that of TDMA only has one peak among the considered range of $M_1$. This indicates that the proposed NOMA has a larger degree of freedom in adaptive rate allocation than that of TDMA under the condition of causing nearly no degradation of system error performance.

\section{Conclusions}

In this paper, we have presented a practical design framework for the non-orthogonal multiple access (NOMA) scheme in a classical two-user multiple access channel (MAC) with quadrature amplitude modulation (QAM) constellations at both users, the sizes of which are not necessarily the same.  More specifically, we aimed to maximize the minimum Euclidean distance of the sum-constellation at the receiver by adjusting the instantaneous transmit power and phase of each user under an individual average power constraint. The design objective was formulated into a \emph{mixed continuous-discrete} optimization problem. By introducing a new mathematical concept termed \emph{punched Farey sequence} and investigating its fundamental properties, we  managed to attain a compact closed-form solution.  Based on the obtained solution, an adaptive rate allocation was performed to further maximize the minimum Euclidean distance of the received sum-constellation under a sum-rate constraint; a high-rate asymptotic approximation solution was also proposed to extract more insights on the NOMA design. Computer simulations were conducted to verify our derivation under various channel configurations, and the simulation results demonstrated that our proposed NOMA scheme outperforms OMA and existing NOMA significantly and the performance gap can be further enlarged when there is a near-far effect between the users.

\section*{Appendix}
\begin{appendices}
\subsection{Proof of Property~\ref{prop:basicFareyprop}}\label{appendix:propbasicFarey}	
1)  We first prove that $\frac{n_1}{m_1}<\frac{n_1 + n_2}{m_1 + m_2}$, which can be showed by calculating
$\frac{n_1 + n_2}{m_1 + m_2}-\frac{n_1}{m_1}=\frac{m_1 n_2 - m_2 n_1}{m_1(m_1+m_2)}>0$,
since $\frac{n_1}{m_1}< \frac{n_2}{m_2}$. The rest cases can be proved in a similar fashion and hence are omitted.

2) We now prove that $m_1 n_2 - m_2 n_1=1$ and it also gives the construction of the term which succeeds $\frac{n_1}{m_1}$ in $\mathfrak{P}_K^L$. First of all, since $\langle  m_1, n_1\rangle=1$, the following equation
\begin{align}\label{eqn:nexttermcondition}
m_1 n - m n_1 =1
\end{align}
has  integer solutions in $m, n$ such that  $m=m_0 + r m_1, n= n_0 + r n_1$ for any integer $r$, where $m_0, n_0$ is a particular set of solutions to~\eqref{eqn:nexttermcondition} and $\langle m, n\rangle=1$~\cite[Thm.\,25]{Hardy75}.
As $\frac{n_1}{m_1}\in \mathfrak{P}_K^L$, we have $0\le n_1 \le L$ and $0\le m_1\le K$. Then, we can choose $m, n$ satisfying either condition:
\begin{subequations}
 \begin{align}
	&{\rm Case~1}: K- m_1 <m \le K, {\rm~and~}  0<n \le L;\label{eqn:case1condition}\\
	&{\rm Case~2}: 0<m \le K-m_1, {\rm~and~}  L-n_1<n \le L.\label{eqn:case2condition}
	\end{align}
\end{subequations}	
Now, since $\frac{n}{m}$ is in its lowest terms (i.e., $\langle  m, n\rangle=1$), and for either case we have $0<m \le K$, $0<n \le L$, we conclude that $\frac{n}{m}$ is a fraction of $\mathfrak{P}_K^L$. In what follows, we will show that either Case 1  or 2 will generate the next term which comes after $\frac{n_1}{m_1}$ in $\mathfrak{P}_K^L$.

{\bf Case 1:} From~\eqref{eqn:nexttermcondition}, $\frac{n}{m}=\frac{n_1}{m_1}+\frac{1}{m_1 m}>\frac{n_1}{m_1}$,
hence $\frac{n}{m}$ comes after $\frac{n_1}{m_1}$ in $\mathfrak{P}_K^L$. Then, if $\frac{n}{m}$ is not $\frac{n_2}{m_2}$, it will come after $\frac{n_2}{m_2}$, and then
	\begin{align}\label{eqn:case1nextterm}
	\frac{n_2}{m_2}-\frac{n_1}{m_1} &= \frac{m_1 n_2 - m_2 n_1}{m_1 m_2} \ge \frac{1}{m_1 m_2};
	~\frac{n}{m}-\frac{n_2}{m_2} =\frac{m_2 n - m n_2}{m_2 m}\ge \frac{1}{m_2 m}.
	\end{align}
As a result, by jointly considering~\eqref{eqn:nexttermcondition} and~\eqref{eqn:case1nextterm}, we have
$\frac{1}{m_1 m}\stackrel{(a)}{=}\frac{m_1 n - m n_1}{m_1 m}= \frac{n}{m}-\frac{n_1}{m_1}
=\frac{n}{m}-\frac{n_2}{m_2}+	\frac{n_2}{m_2}-\frac{n_1}{m_1} \stackrel{(b)}{\ge}   \frac{1}{m_2 m}+\frac{1}{m_1 m_2}
=\frac{m_1+m}{m_1 m_2 m}\stackrel{(c)}{>} \frac{K}{m_1 m_2 m} \stackrel{(d)}{\ge}   \frac{1}{m_1 m}$,
where $(a)$ follows from~\eqref{eqn:nexttermcondition}; inequality $(b)$ holds since~\eqref{eqn:case1nextterm}; inequality $(c)$ follows from~\eqref{eqn:case1condition} and $(d)$ is true since $\frac{n_2}{m_2}\in \mathfrak{P}_K^L$.
This is a contradiction, and therefore $\frac{n}{m}$ must be $\frac{n_2}{m_2}$, and hence $m_1 n_2 - m_2 n_1=1$.

{\bf Case 2:} As in Case 1, according to~\eqref{eqn:nexttermcondition}, $\frac{n}{m}=\frac{n_1}{m_1}+\frac{1}{m_1 m}>\frac{n_1}{m_1}$, and hence $\frac{n}{m}$ comes after $\frac{n_1}{m_1}$ in $\mathfrak{P}_K^L$. As a result, if $\frac{n}{m}$ is not $\frac{n_2}{m_2}$, it comes after $\frac{n_2}{m_2}$, and we have
	\begin{align}\label{eqn:case2nextterm}
	\frac{m_1}{n_1}-\frac{m_2}{n_2} = \frac{m_1 n_2-m_2 n_1 }{n_1 n_2} \ge \frac{1}{n_1 n_2};~
	\frac{m_2}{n_2}-\frac{m}{n} =\frac{m_2 n-m n_2}{n_2 n} \ge \frac{1}{n_2 n}.
	\end{align}
Likewise, we conclude that
	$\frac{1}{n_1 n}\stackrel{(a)}{=}\frac{m_1 n - m n_1}{n_1 n}=\frac{m_1}{n_1}-\frac{m}{n}
	=\frac{m_1}{n_1}-\frac{m_2}{n_2}+ \frac{m_2}{n_2}-\frac{m}{n} \stackrel{(b)}{\ge}  \frac{1}{n_1 n_2}+ \frac{1}{n_2 n}
	= \frac{n+ n_1}{n_1 n_2 n} \stackrel{(c)}{>}\frac{L}{n_1 n_2 n} \stackrel{(d)}{\ge} \frac{1}{n_1 n}$,
where $(a)$ follows from~\eqref{eqn:nexttermcondition}, inequality $(b)$ holds as a result of~\eqref{eqn:case2nextterm}, and inequality $(c)$ follows from~\eqref{eqn:case2condition}, and $(d)$  is true since $\frac{n_2}{m_2}\in \mathfrak{P}_K^L$. There is also a contradiction,  therefore $\frac{n}{m}$ must be $\frac{n_2}{m_2}$, and thus $m_1 n_2 - m_2 n_1=1$.
3) From the above proof, i.e.,~\eqref{eqn:case1condition} and~\eqref{eqn:case2condition}, we can observe that: (a) If $n_1+n_2\le L$, then $m_1+m_2>K$; (b) If $m_1+m_2\le K$, then $n_1+n_2>L$.

4) {From~\eqref{eqn:case1condition} and~\eqref{eqn:case2condition}, we can observe that, $m_1+m_2>0$ (i.e., $m_1+m_2 \ge 1$) and $n_1+n_2>0$ (i.e., $n_1+n_2 \ge 1$).
First, we consider the case $\frac{n_1}{m_1}=\frac{0}{1}$. As $n_1=0$, then by solving~\eqref{eqn:nexttermcondition} of Case 1 in the above discussion, we have $n_2=1$. By~\eqref{eqn:case1condition}, we attain $K- 1 <m_2 \le K$, i.e., $m_2=K$. Now, we attain two adjacent terms $\frac{n_1}{m_1}=\frac{0}{1}$ and $\frac{n_2}{m_2}=\frac{1}{K}$ such that $n_1+n_2=1$ and $m_1+m_2=K+1$. Similarly, we can find adjacent terms $\frac{n_1}{m_1}=\frac{L}{1}$ an $\frac{n_2}{m_2}=\frac{1}{0}$ such that $m_1+m_2=1$ and $n_1+n_2=L+1$.
}

This completes the proof of Property~\ref{prop:basicFareyprop}.~\hfill$\Box$
	
\subsection{Proof of Property~\ref{prop:mediumvalue}}\label{appendix:mediumvalue}		
By Property~\ref{prop:basicFareyprop}, we have
$m_1 n_2 -m_2 n_1=1$,
$m_2 n_3 -m_3 n_2=1$.
Then, solving the following equations:
$m_3 m_1 n_2 - m_3 m_2 n_1 =m_3$, $m_1 m_2 n_3 - m_1 m_3 n_2=m_1$,
$n_3 m_1 n_2 - n_3 m_2 n_1=n_3$, $n_1 m_2 n_3 - n_1 m_3 n_2=n_1$,
for $m_2$, $n_2$, we attain
$m_2 (m_1 n_3 - m_3 n_1) =m_1 + m_3$,
$n_2(m_1 n_3 -m_3 n_1)  =n_1 + n_3$.
As $m_1 n_3 - m_3 n_1 \neq 0$, we have $\frac{n_2}{m_2}=\frac{n_1+n_3}{m_1+m_3}$. The property is proved.
~\hfill$\Box$
\subsection{Proof of Property~\ref{prop:inequalitymodified}}\label{appendix:inequalitymodified}
From the assumption, we have
	$\frac{n_2}{m_2} -\frac{n_1+n_3}{m_1+m_3}=\frac{m_1 n_2 + m_3 n_2 - m_2 n_1 -m_2 n_3}{m_2(m_1+m_3)}
	=\frac{m_1 n_2 - m_2 n_1 -1}{m_2(m_1+m_2)}\ge 0$;
	$\frac{n_2+n_4}{m_2+m_4}-\frac{n_3}{m_3}=\frac{m_3 n_2 + m_3 n_4 - m_2 n_3 -m_4 n_3}{m_3(m_2+m_4)}
	=\frac{m_3 n_4 - m_4 n_3-1}{m_3(m_2+m_4)}\ge 0$.
The completes the proof.~\hfill$\square$

\subsection{Proof of Proposition\,\ref{prop:basicfareyextended}}\label{appendix:fareydiv}
Recall that $d(m,n)=\big||\tilde{h}_1| \tilde{w}_1 n-|\tilde{h}_2| \tilde{w}_2 m\big|$.  Therefore, for $\frac{|\tilde{h}_2| \tilde{w}_2}{|\tilde{h}_1 |\tilde{w}_1}  \in \big(\frac{n_1}{m_1},\frac{n_2}{m_2}\big)$, we have
$d(m_1, n_1)-d(m_2, n_2)
=\big||\tilde{h}_1|\tilde{w}_1 n_1 -|\tilde{h}_2| \tilde{w}_2 m_1\big| -\big||\tilde{h}_1|\tilde{w}_1 n_2 - |\tilde{h}_2| \tilde{w}_2 m_2\big|
=-|\tilde{h}_1|\tilde{w}_1 n_1 + |\tilde{h}_2| \tilde{w}_2 m_1 - |\tilde{h}_1|\tilde{w}_1 n_2 + |\tilde{h}_2| \tilde{w}_2 m_2
=  (m_1+m_2) |\tilde{h}_1|\tilde{w}_1 \Big(\frac{|\tilde{h}_2| \tilde{w}_2}{ |\tilde{h}_1| \tilde{w}_1} - \frac{n_1+n_2}{ m_1+m_2}\Big)$.
The results in Proposition\,\ref{prop:basicfareyextended} can be readily obtained, and we complete the proof.
\hfill$\square$
\subsection{Proof of of Proposition\,\ref{prop:worstcasemodified}}\label{appendix:worstcase}
\emph{Proof:} As $\frac{n_1}{m_1}$ and $\frac{n_4}{m_4}$ are arbitrarily chosen, Proposition\,\ref{prop:worstcasemodified} is equivalent to
\begin{enumerate}
	\item If $\frac{|\tilde{h}_2| \tilde{w}_2}{|\tilde{h}_1|\tilde{w}_1} \in \big(\frac{n_2}{m_2}, \frac{n_2 +n_3}{m_2 + m_3}\big)$, then $d(m_2, n_2) <d(m_3, n_3)$, $d(m_2, n_2) <d(m_4, n_4)$, and $d(m_2, n_2) <d(m_1, n_1)$;
	\item If  $\frac{|\tilde{h}_2| \tilde{w}_2}{|\tilde{h}_1|\tilde{w}_1} \in \big(\frac{n_2 +n_3}{m_2 + m_3}, \frac{n_3}{m_3}\big)$, then $d(m_3, n_3) <d(m_2, n_2)$, $d(m_3, n_3) <d(m_1, n_1)$, and $d(m_3, n_3) <d(m_4, n_4)$.
\end{enumerate}

First, by Proposition~\ref{prop:basicfareyextended}, we have
\begin{enumerate}
	\item If $\frac{|\tilde{h}_2| \tilde{w}_2}{|\tilde{h}_1|\tilde{w}_1} \in \big(\frac{n_2}{m_2}, \frac{n_2 +n_3}{m_2 + m_3}\big)$, then $d(m_2, n_2) <d(m_3, n_3)$ and $d(m_2, n_2) <d(m_4, n_4)$;
	\item If  $\frac{|\tilde{h}_2| \tilde{w}_2}{|\tilde{h}_1| \tilde{w}_1} \in \big(\frac{n_2 +n_3}{m_2 + m_3}, \frac{n_3}{m_3}\big)$, then $d(m_3, n_3) <d(m_1, n_1)$ and $d(m_3, n_3) <d(m_2, n_2)$.
\end{enumerate}

Then, we want to show that:
\begin{enumerate}
	\item If $\frac{|\tilde{h}_2| \tilde{w}_2}{|\tilde{h}_1| \tilde{w}_1} \in \big(\frac{n_2}{m_2}, \frac{n_2 +n_3}{m_2 + m_3}\big)$, then $d(m_2, n_2) <d(m_1, n_1)$;
	\item If  $\frac{|\tilde{h}_2| \tilde{w}_2}{|\tilde{h}_1| \tilde{w}_1} \in \big(\frac{n_2 +n_3}{m_2 + m_3}, \frac{n_3}{m_3}\big)$, then $d(m_3, n_3)<d(m_4, n_4)$.
\end{enumerate}
The first case can be proved by considering
	$d(m_1, n_1)- d(m_2, n_2)=\big||\tilde{h}_1| \tilde{w}_1 n_1 -|\tilde{h}_2| \tilde{w}_2 m_1\big| -\big||\tilde{h}_1|\tilde{w}_1 n_2 - |\tilde{h}_2| \tilde{w}_2 m_2\big|
	=   |\tilde{h}_1|\tilde{w}_1 \Big(\frac{| \tilde{h}_2| w_2}{ |\tilde{h}_1|w_1} (m_1 -m_2) - (n_1-n_2)\Big)$.
As $\frac{|\tilde{h}_2| \tilde{w}_2}{|\tilde{h}_1| \tilde{w}_1} \in \big(\frac{n_2}{m_2}, \frac{n_2 +n_3}{m_2 + m_3}\big)$, then $d(m_1, n_1)- d(m_2, n_2)\ge 0$ is true if $\frac{n_2}{m_2} (m_1 -m_2) - (n_1-n_2)\ge 0$ and $\frac{n_2 +n_3}{m_2 + m_3} (m_1 -m_2) - (n_1-n_2)\ge 0$.
We know that $\frac{n_2}{m_2} (m_1 -m_2) - (n_1-n_2)
	=\big( m_1 n_2 - m_2n_1\big)/m_2 > 0$,
and $\frac{n_2 +n_3}{m_2 + m_3} (m_1 -m_2) - (n_1-n_2)
	=\big((n_2 +n_3) (m_1 -m_2) - (m_2 + m_3)(n_1-n_2)\big)/(m_2+m_3)
	= \big((m_1 + m_3)n_2- m_2(n_1+n_3) +m_1 n_3 - m_3 n_1\big)/(m_2+m_3)>0$,
where the inequality is always true by Property~\ref{prop:inequalitymodified}.

Likewise, the second case can be proved by considering
      $d(m_4, n_4)- d(m_3, n_3) =\big|| \tilde{h}_1|\tilde{w}_1 n_4 -|\tilde{h}_2| \tilde{w}_2 m_4\big| -\big|| \tilde{h}_1|\tilde{w}_1 n_3 - | \tilde{h}_2| \tilde{w}_2 m_3\big|
	=   |\tilde{h}_1|\tilde{w}_1 \Big(\frac{|\tilde{h}_2| \tilde{w}_2}{ |\tilde{h}_1| \tilde{w}_1} (m_3 -m_4) - (n_3-n_4)\Big)$.
As $\frac{|\tilde{h}_2| \tilde{w}_2}{|\tilde{h}_1|\tilde{w}_1} \in \big(\frac{n_2 +n_3}{m_2 + m_3}, \frac{n_3}{m_3}\big)$, then $d(m_4, n_4)- d(m_3, n_3)\ge 0$ is true if $\frac{n_3}{m_3} (m_3 -m_4) - (n_3-n_4)\ge 0$ and $\frac{n_2 +n_3}{m_2 + m_3} (m_3 -m_4) - (n_3-n_4)\ge 0$. We know that $\frac{n_3}{m_3} (m_3 -m_4)-(n_3-n_4)=\big(n_3(m_3 -m_4) - m_3(n_3-n_4)\big)/m_3= (m_3 n_4- m_4 n_3)/m_3>0$,
and $\frac{n_2 +n_3}{m_2 + m_3} (m_3-m_4) - (n_3-n_4)
	= \big((n_2 +n_3) (m_3 -m_4) - (m_2 + m_3)(n_3-n_4)\big)/(m_2+m_3)
	= \big(m_3(n_2 + n_4) - n_3(m_2+m_4) + m_2 n_4 - m_4 n_2\big)/(m_2+m_3)>0$,
where the inequality is always true by Property~\ref{prop:inequalitymodified}. We complete the proof.\hfill$\Box$

\subsection{Proof of Lemma\,\ref{thm:subinterval}} \label{appendix:theorem1}

According to proposition~\ref{prop:worstcasemodified} and notice that $\big(\frac{b_k}{a_k},\frac{b_{k+1}}{a_{k+1}}\big)=\big(\frac{b_k}{a_k}, \frac{b_k+b_{k+1}}{a_k+a_{k+1}}\big)\cup \big(\frac{b_k+b_{k+1}}{a_k+a_{k+1}},\frac{b_{k+1}}{a_{k+1}} \big)$, problem in~\eqref{eqn:maxminbyinterval} can be further divided into the following two sub-problems, and the overall solution is the maximum value of the two problems:
\begin{problem}[Sub-problem 1]\label{pbm:subpbm1}
		The optimization problem is stated as follows:
	\begin{subequations}\label{eqn:subpbm1}
		\begin{align}
		&g_1\Big(\frac{b_k}{a_k}, \frac{b_{k+1}}{a_{k+1}}\Big)=\max_{(\tilde{w}_1, \tilde{w}_2)}~|\tilde{h}_2| \tilde{w}_2 a_k- | \tilde{h}_1| \tilde{w}_1 b_k\\
		&{\rm s.t.~}\frac{b_k}{a_k}\le \frac{|\tilde{h}_2|\tilde{w}_2 }{|\tilde{h}_1| \tilde{w}_1} < \frac{b_k+b_{k+1}}{a_k+a_{k+1}},~ 0 < \tilde{w}_1 \le 1,~0 < \tilde{w}_2 \le 1.
		\end{align}
		\hfill$\blacksquare$
	\end{subequations}
\end{problem}
\begin{problem}[Sub-problem 2]\label{pbm:subpbm2}
We aim to solve the following optimization problem:
\begin{subequations}\label{eqn:subpbm2}
	\begin{align}
	&g_2\Big(\frac{b_k}{a_k}, \frac{b_{k+1}}{a_{k+1}}\Big)=\max_{(\tilde{w}_1, \tilde{w}_2)}~|\tilde{h}_1| \tilde{w}_1 b_{k+1} - |\tilde{h}_2| \tilde{w}_2 a_{k+1}\\
	&{\rm s.t.~}\frac{b_k+b_{k+1}}{a_k+a_{k+1}} \le \frac{|\tilde{h}_2| \tilde{w}_2 }{|\tilde{h}_1| \tilde{w}_1}\le \frac{b_{k+1}}{a_{k+1}},~ 0< \tilde{w}_1 \le 1, 0 < \tilde{w}_2 \le 1.
	\end{align}
	\hfill$\blacksquare$
\end{subequations}		
\end{problem}
We know that~\eqref{eqn:subpbm1} is equivalent to
\begin{subequations}
	\begin{align}
 &g_1\Big(\frac{b_k}{a_k}, \frac{b_{k+1}}{a_{k+1}}\Big)=\max_{(w_1, w_2)}~|\tilde{h}_2|\tilde{w}_2 a_k- |\tilde{h}_1| \tilde{w}_1 b_k\\
 &{\rm s.t.~}  \frac{|\tilde{h}_2| (a_k +a_{k+1}) }{|\tilde{h}_1| (b_k+b_{k+1})} \tilde{w}_2< \tilde{w}_1 \le \frac{a_k|\tilde{h}_2|}{b_k |\tilde{h}_1|} \tilde{w}_2,~ 0< \tilde{w}_1 \le 1,~0< \tilde{w}_2 \le 1.
	\end{align}
\end{subequations}

We can find that the objective function is a linear decreasing function of $\tilde{w}_1$. Then, we let $\tilde{w}_1= \frac{|\tilde{h}_2| (a_k +a_{k+1}) }{|\tilde{h}_1| (b_k+b_{k+1})} \tilde{w}_2$, and the objective function can be reformulated by
	$|\tilde{h}_2|\tilde{w}_2 a_k- |\tilde{h}_1| \tilde{w}_1 b_k
	=\big(a_k(b_k+b_{k+1})-b_k  (a_k +a_{k+1}) \big) \frac{|\tilde{h}_2|\tilde{w}_2}{b_k+b_{k+1}}
	\stackrel{(a)}{=}\frac{|\tilde{h}_2|\tilde{w}_2}{b_k+b_{k+1}}$,
where $(a)$ follows from Property~\ref{prop:basicFareyprop}. Now, the constraints on $\tilde{w}_2$ are
	$0< \tilde{w}_2 \le 1,~0< \tilde{w}_2 \le \frac{|\tilde{h}_1| (b_k+b_{k+1})}{|\tilde{h}_2|(a_k+a_{k+1})}$.
Therefore, the solution to~\eqref{eqn:subpbm1} can be given as follows:
    \begin{align}
g_1\Big(\frac{b_k}{a_k}, \frac{b_{k+1}}{a_{k+1}}\Big)=\begin{cases}
            \frac{|\tilde{h}_2|}{b_k+b_{k+1}},~{\rm with~} (\tilde{w}_1, \tilde{w}_2)=(\frac{|\tilde{h}_2| (a_k   +a_{k+1}) }{|\tilde{h}_1| (b_k+b_{k+1})},1), &  {\rm if~}\frac{|\tilde{h}_2|}{|\tilde{h}_1|} \le \frac{b_k+b_{k+1}}{a_k+a_{k+1}};\\
            \frac{|\tilde{h}_1| }{a_k+a_{k+1}},~{\rm with~} (\tilde{w}_1, \tilde{w}_2)=(1, \frac{|\tilde{h}_1| (b_k+b_{k+1})}{|\tilde{h}_2|(a_k+a_{k+1})}),
            &  {\rm if~}\frac{|\tilde{h}_2|}{|\tilde{h}_1|} > \frac{b_k+b_{k+1}}{ a_k+a_{k+1}}.
            \end{cases}
	\end{align}

Likewise, we note that~\eqref{eqn:subpbm2} is equivalent to
\begin{subequations}
	\begin{align}
	&g_2\Big(\frac{b_k}{a_k}, \frac{b_{k+1}}{a_{k+1}}\Big)=\max_{(\tilde{w}_1, \tilde{w}_2)}~|\tilde{h}_1| \tilde{w}_1 b_{k+1} - |\tilde{h}_2|\tilde{w}_2 a_{k+1}\\
	&{\rm s.t.~}\frac{(b_k+b_{k+1})|\tilde{h}_1|}{(a_k+a_{k+1})|\tilde{h}_2|} \tilde{w}_1 \le \tilde{w}_2 \le \frac{b_{k+1} |\tilde{h}_1|}{a_{k+1} |\tilde{h}_2|} \tilde{w}_1,~ 0< \tilde{w}_1 \le 1, 0< \tilde{w}_2 \le 1.
	\end{align}
\end{subequations}	
By letting $\tilde{w}_2=\frac{(b_k+b_{k+1})|\tilde{h}_1|}{(a_k+a_{k+1})|\tilde{h}_2|} \tilde{w}_1$, the objective function can be reformulated by
$|\tilde{h}_1| \tilde{w}_1 b_{k+1} - |\tilde{h}_2|\tilde{w}_2 a_{k+1}
=\big(b_{k+1}(a_k+a_{k+1}) - a_{k+1} (b_k+b_{k+1})\big)\frac{|\tilde{h}_1|\tilde{w}_1}{a_k+a_{k+1}}
=\frac{|\tilde{h}_1| \tilde{w}_1}{a_k+a_{k+1}}$.
The constraints on $\tilde{w}_1$ are $0< \tilde{w}_1 \le 1$ and $0< \tilde{w}_1 < \frac{(a_k+a_{k+1})|h_2|}{(b_k+b_{k+1})|\tilde{h}_1|}$. Thus, we have
    \begin{align}
g_2 \Big(\frac{b_k}{a_k}, \frac{b_{k+1}}{a_{k+1}}\Big)=\begin{cases}
            \frac{|\tilde{h}_2|}{b_k+b_{k+1}}, ~{\rm with~} (\tilde{w}_1, \tilde{w}_2)=(\frac{|\tilde{h}_2| (a_k   +a_{k+1}) }{|\tilde{h}_1| (b_k+b_{k+1})},1),  & {\rm if~}  \frac{|\tilde{h}_2|}{|\tilde{h}_1|} \le \frac{b_k+b_{k+1}}{a_k+a_{k+1}};\\
            \frac{|\tilde{h}_1| }{a_k+a_{k+1}}{\rm~with~} (\tilde{w}_1, \tilde{w}_2)=(1, \frac{|\tilde{h}_1| (b_k+b_{k+1})}{|\tilde{h}_2|(a_k+a_{k+1})}),
            & {\rm if~} \frac{|\tilde{h}_2|}{|\tilde{h}_1|} > \frac{b_k+b_{k+1}}{ a_k+a_{k+1}}.
            \end{cases}
	\end{align}
Combining the two cases,  we have the result in Lemma~\ref{thm:subinterval}, and we complete the proof.\hfill$\Box$

\subsection{Proof of Theorem~\ref{thm:gaussianmacpower}}\label{appendix:theoremac}
Consider the punched Farey sequence $\mathfrak{P}_{M_2-1}^{M_1-1}=\big(\frac{b_1}{a_1},\frac{b_2}{a_2},\cdots, \frac{b_{C}}{a_{C}}\big)$, where $C=|\mathfrak{P}_{M_2-1}^{M_1-1}|$. We consider each case separately as follows:

 1) If $\frac{|\tilde{h}_2|}{|\tilde{h}_1|} \le \frac{1}{M_2}$, we have $\frac{|\tilde{h}_2|}{|\tilde{h}_1|} \le \frac{1}{M_2}=\frac{b_1+b_2}{a_1+a_2} \le \frac{b_k+b_{k+1}}{ a_k+a_{k+1}}, k=1, \ldots, C-1$. By Lemma~\ref{thm:subinterval}, for each Farey interval, we can attain that
	$g\Big(\frac{b_{k}}{a_{k}}, \frac{b_{k+1}}{a_{k+1}}\Big)= \frac{|\tilde{h}_2| }{b_k+b_{k+1}}$, for $k=1, \ldots, C-1$.
	As a consequence, the minimum Euclidean distance $d^*$ can be attained by taking the maximum value of the objective function over all the possible intervals, given by:
	$d^* =\max \Big\{\frac{|\tilde{h}_2| }{b_1+b_2}, \ldots, \frac{|\tilde{h}_2| }{b_{C-1}+b_C}\Big\}
	=\frac{|\tilde{h}_2| }{b_1+b_2}=|\tilde{h}_2|$,
	where the optimality is attained when $(\tilde{w}_1^*, \tilde{w}_2^*)= (M_2 \frac{|\tilde{h}_2|}{|\tilde{h}_1|},1)$ with the help of Property~\ref{prop:basicFareyprop}, and hence $\frac{|\tilde{h}_1| \tilde{w}_1^*}{|\tilde{h}_2| \tilde{w}_2^*}=M_2$.
	
 2) If $\frac{1}{M_2}<\frac{|\tilde{h}_2|}{|\tilde{h}_1|} \le \frac{M_1}{M_2}$, we can suppose that $\frac{b_{\ell_1}+b_{\ell_1+1}}{ a_{\ell_1}+a_{\ell_1+1}} < \frac{|\tilde{h}_2|}{|\tilde{h}_1|} \le \frac{b_{\ell_1+1}+b_{\ell_1+2}}{a_{\ell_1+1}+a_{\ell_1+2}}$, where $\ell_1$ can be determined upon  the knowledge of $\frac{|\tilde{h}_2|}{|\tilde{h}_1|}$. Then, with the help of Lemma~\ref{thm:subinterval}, we have
	\begin{align*}
	g\Big(\frac{b_{k}}{a_{k}}, \frac{b_{k+1}}{a_{k+1}}\Big)=\begin{cases}
	\frac{|\tilde{h}_1|}{a_k+a_{k+1}}, ~k=1, \ldots, \ell_1; \\
	\frac{|\tilde{h}_2|}{b_k+b_{k+1}}, ~ k=\ell_1+1, \ldots, C-1.
	\end{cases}
	\end{align*}
	First, for $a_k+a_{k+1}, k=1, \ldots, \ell_1$, we have the following two cases:
		(a) If $a_k+a_{k+1}\ge M_2$, then we have $\frac{1}{a_k+a_{k+1}} \le \frac{1}{M_2}$;
		(b) If $a_k+a_{k+1}< M_2$ (i.e., $a_k+a_{k+1}\le  M_2-1$), then by~Property~\ref{prop:basicFareyprop}, we have $b_k+b_{k+1} \ge M_1$ (i.e., $b_k+b_{k+1} > M_1-1$). From the assumption, we have
		$\frac{b_k+b_{k+1}}{a_k+a_{k+1}} \le \frac{b_{\ell_1}+b_{\ell_1+1}}{ a_{\ell_1}+a_{\ell_1+1}} < \frac{|\tilde{h}_2|}{|\tilde{h}_1|} \le \frac{M_1}{M_2}$. Therefore, we have $\frac{1}{a_k+a_{k+1}} \le \frac{M_1}{M_2(b_k+b_{k+1})} \le \frac{1}{M_2}$.
	Combining the above two cases, we have
	\begin{align}\label{eqn:mindencase1}
	\frac{1}{a_k+a_{k+1}} \le \frac{1}{M_2}, ~ k=1, \ldots, \ell_1.
	\end{align}
	
	Next, consider $b_k+b_{k+1}, k=\ell_1+1, \ldots, C-1$ and we can show that:
		(a) If $b_k+b_{k+1} < M_1$ (i.e., $b_k+b_{k+1} \le M_1-1$), then by~Property~\ref{prop:basicFareyprop}, we have $a_k+a_{k+1}\ge M_2$ (i.e., $a_k+a_{k+1}> M_2-1$). As a consequence, we have $\frac{|\tilde{h}_2|}{|\tilde{h}_1|} \le \frac{b_{\ell_1+1}+b_{\ell_1+2}}{a_{\ell_1+1}+a_{\ell_1+2}} \le \frac{b_k+b_{k+1}}{a_k+a_{k+1}} \le \frac{b_k+b_{k+1}}{M_2}$;
		(b) If $b_k+b_{k+1} \ge M_1$, then we have $\frac{|\tilde{h}_2|}{|\tilde{h}_1|} \le  \frac{M_1}{M_2}  \le \frac{b_k+b_{k+1}}{M_2}$.
	Combining both cases, we have
	\begin{align}\label{eqn:minnumcase1}
	\frac{|\tilde{h}_1|}{M_2}\ge \frac{|\tilde{h}_2|}{b_k+b_{k+1}},  k=\ell_1+1, \ldots, C-1.
	\end{align}
	Now, with the help of~\eqref{eqn:mindencase1} and \eqref{eqn:minnumcase1}, the overall minimum Euclidean distance is given by
		$d^*=\max~\big\{
		\frac{|\tilde{h}_1|}{a_1+a_2} , \ldots, \frac{|\tilde{h}_1|}{a_{\ell_1}+a_{\ell_1+1}}, \frac{|\tilde{h}_2|}{b_{\ell_1+1}+b_{\ell_1+2}}, \ldots,  \frac{|\tilde{h}_2|}{b_{C-1}+b_{C}}\big\}
		=\max \big\{\frac{|\tilde{h}_1|}{M_2}, \frac{|\tilde{h}_2|}{b_{\ell_1+1}+b_{\ell_1+2}}, \ldots,  \frac{|\tilde{h}_2|}{b_{C-1}+b_{C}} \big\}=\frac{|\tilde{h}_1|}{M_2}$,
	where the optimality is attained when $(\tilde{w}_1^*, \tilde{w}_2^*)= (1, \frac{|\tilde{h}_1|}{M_2 |\tilde{h}_2|})$ and as a result we have $\frac{|\tilde{h}_1| \tilde{w}_1^*}{|\tilde{h}_2| \tilde{w}_2^*}=M_2$.
	
 3) If $\frac{M_1}{M_2}<\frac{|\tilde{h}_2|}{|\tilde{h}_1|} \le M_1$, we can suppose that $\frac{b_{\ell_2}+b_{\ell_2+1}}{ a_{\ell_2}+a_{\ell_2+1}} \le \frac{|\tilde{h}_2|}{|\tilde{h}_1|} <\frac{b_{\ell_2+1}+b_{\ell_2+2}}{ a_{\ell_2+1}+a_{\ell_2+2}}$. With the help of Lemma~\ref{thm:subinterval}, we have
	\begin{align*}
	g\Big(\frac{b_{k}}{a_{k}}, \frac{b_{k+1}}{a_{k+1}}\Big)=\begin{cases}
	\frac{|\tilde{h}_1|}{a_k+a_{k+1}}, ~k=1, \ldots, \ell_2;\\
	\frac{|\tilde{h}_2|}{b_k+b_{k+1}}, ~k=\ell_2+1, \ldots, C-1.
	\end{cases}
	\end{align*}
	
	We first show that, for $b_k+b_{k+1}, k=\ell_2+1, \ldots, C-1$,
		(a) If $b_k+b_{k+1}\ge M_1$, then we have $\frac{1}{b_k+b_{k+1}} \le \frac{1}{M_1}$;
		(b) If $b_k+b_{k+1}< M_1$, then by~Property~\ref{prop:basicFareyprop}, we have $a_k+a_{k+1} \ge M_2$. From the assumption, we have
		$\frac{b_k+b_{k+1}}{a_k+a_{k+1}} \ge \frac{b_{\ell_2+1}+b_{\ell_2+2}}{ a_{\ell_2+1}+a_{\ell_2+2}} > \frac{|\tilde{h}_2|}{|\tilde{h}_1|} > \frac{M_1}{M_2}$. Therefore, we have $\frac{1}{b_k+b_{k+1}} < \frac{M_2}{M_1(a_k+a_{k+1})} \le \frac{1}{M_1}$.
	By jointly considering both cases, we have
	\begin{align}\label{eqn:mindencase2}
	\frac{1}{b_k+b_{k+1}} \le \frac{1}{M_1}, {~\rm for~}k=\ell_2+1, \ldots, C-1.
	\end{align}
	Next, we consider $a_k+a_{k+1}, k=1, \ldots, \ell_2$,
		(a) If $a_k+a_{k+1} < M_2$, then by~Property~\ref{prop:basicFareyprop}, we have $b_k+b_{k+1} \ge M_1$;
		As a result, $\frac{M_1}{a_k+a_{k+1}} \le \frac{b_k+b_{k+1}}{a_k+a_{k+1}} \le  \frac{b_{\ell_2}+b_{\ell_2+1}}{ a_{\ell_2}+a_{\ell_2+1}} \le \frac{|\tilde{h}_2|}{|\tilde{h}_1|}$.
		(b) If $a_k+a_{k+1}\ge M_2$, then $\frac{M_1}{a_k+a_{k+1}} \le \frac{M_1}{M_2} <\frac{|\tilde{h}_2|}{|\tilde{h}_1|}$.
	Combining both cases, we conclude that
	\begin{align}\label{eqn:minnumcase2}
	\frac{|\tilde{h}_1|}{a_k+a_{k+1}} \le \frac{|\tilde{h}_2|}{M_1}, k=1, \ldots, \ell_2.
	\end{align}
	
Therefore, with the help of~\eqref{eqn:mindencase2} and~\eqref{eqn:minnumcase2}, the overall minimum Euclidean distance is
		$d^*=\max~\big\{
		\frac{|\tilde{h}_1|}{a_1+a_2}, \ldots, \frac{|\tilde{h}_1|}{a_{\ell_2}+a_{\ell_2+1}}, \frac{|\tilde{h}_2|}{b_{\ell_2+1}+b_{\ell_2+2}}, \ldots, \frac{|\tilde{h}_2|}{b_{C-1}+b_C}\big\}
		=\max~\big\{\frac{|\tilde{h}_1|}{a_1+a_2}, \ldots, \frac{|\tilde{h}_1|}{a_{\ell_2}+a_{\ell_2+1}}, \frac{|\tilde{h}_2|}{M_1}\big\}
		=\frac{|\tilde{h}_2|}{M_1}$,
	where the optimality is attained when $(\tilde{w}_1^*, \tilde{w}_2^*)=\big(\frac{|\tilde{h}_2| }{M_1 |\tilde{h}_1|}, 1\big)$ and as a result, $d^*=\frac{|\tilde{h}_2|}{M_1}$ and $\frac{|\tilde{h}_1| \tilde{w}_1^*}{|\tilde{h}_2| \tilde{w}_2^*}=\frac{1}{M_1}$.
	
 4) If $M_1<\frac{|\tilde{h}_2|}{|\tilde{h}_1|}$, then $\frac{b_k+b_{k+1}}{ a_k+a_{k+1}}\le M_1 <\frac{|\tilde{h}_2|}{|\tilde{h}_1|}$, for $k=1,\ldots, C-1$. By using Lemma~\ref{thm:subinterval},  $g\big(\frac{b_{k}}{a_{k}}, \frac{b_{k+1}}{a_{k+1}}\big)= \frac{|\tilde{h}_1| }{a_k+a_{k+1}}$ for $k=1,\ldots, C-1$, and
	$d^* =\max \Big\{\frac{|\tilde{h}_1| }{a_1+a_2}, \ldots, \frac{|\tilde{h}_1|}{a_{C-1}+a_C}\Big\}
	=\frac{|\tilde{h}_1|}{a_{C-1}+a_C}=|\tilde{h}_1|$,
	where the optimality is attained when $(\tilde{w}_1, \tilde{w}_2)= (1, M_1 \frac{|\tilde{h}_1|}{|\tilde{h}_2|})$ with the help of Property~\ref{prop:basicFareyprop}, and as a result, $\frac{|\tilde{h}_1| \tilde{w}_1^*}{|\tilde{h}_2| \tilde{w}_2^*}=\frac{1}{M_1}$.

The solution to Problem~\ref{pbm:maxminopt} can be summarized as
	\begin{itemize}
		\item If $\frac{|\tilde{h}_2|}{|\tilde{h}_1|} \le \frac{1}{M_2}$, then $(\tilde{w}_1^*, \tilde{w}_2^*)=\big(M_2 \frac{|\tilde{h}_2| }{|\tilde{h}_1|}, 1\big)$, $d^*=|\tilde{h}_2|$, and $\frac{|\tilde{h}_1| \tilde{w}_1^*}{|\tilde{h}_2| \tilde{w}_2^*}=M_2$;
		\item If $\frac{1}{M_2}<\frac{|\tilde{h}_2|}{|\tilde{h}_1|} \le \frac{M_1}{M_2}$, then $(\tilde{w}_1^*, \tilde{w}_2^*)=\big(1, \frac{|\tilde{h}_1| }{M_2 |\tilde{h}_2|}\big)$, $d^*=\frac{|\tilde{h}_1|}{M_2}$, and $\frac{|\tilde{h}_1| \tilde{w}_1^*}{|\tilde{h}_2| \tilde{w}_2^*}=M_2$;
		\item If $\frac{M_1}{M_2}<\frac{|\tilde{h}_2|}{|\tilde{h}_1|} \le M_1$, then $(\tilde{w}_1^*, \tilde{w}_2^*)=\big(\frac{|\tilde{h}_2| }{M_1 |\tilde{h}_1|}, 1\big)$, $d^*=\frac{|\tilde{h}_2|}{M_1}$, and $\frac{|\tilde{h}_1| \tilde{w}_1^*}{|\tilde{h}_2| \tilde{w}_2^*}=\frac{1}{M_1}$;
		\item If $M_1 <\frac{|\tilde{h}_2|}{|\tilde{h}_1|}$, then $(\tilde{w}_1^*, \tilde{w}_2^*)=\big(1, M_1\frac{|\tilde{h}_1| }{|\tilde{h}_2|}\big)$, $d^*=|\tilde{h}_1|$, and $\frac{|\tilde{h}_1| \tilde{w}_1^*}{|\tilde{h}_2| \tilde{w}_2^*}=\frac{1}{M_1}$.
	\end{itemize}
From the previous assumption, we know that $\tilde{w}_1=\sqrt{\frac{2(M_1^2-1)}{3 P_1}} w_1$, $\tilde{w}_2=\sqrt{\frac{2(M_2^2-1)}{3 P_2}} w_2$, $|\tilde{h}_1|=\sqrt{\frac{3 P_1}{2(M_1^2-1)}}|h_1|$, and $|\tilde{h}_2| =\sqrt{\frac{3 P_2}{2(M_2^2-1)}}|h_2|$.  After some algebraic manipulations, the conclusion in Theorem\,\ref{thm:gaussianmacpower} can be readily obtained and we complete the proof of the theorem.~\hfill$\Box$

\subsection{Proof of Corollary~\ref{cor:uniformconst}}\label{appendix:cormindist}
Without loss of generality, we consider $\frac{|h_2|}{|h_1|} \le\sqrt{\frac{P_1 (M_2^2-1)}{P_2 M_2^2(M_1^2-1)}}$, and therefore $|h_1| w_1^* s_1 +|h_2| w_2^* s_2 = \sqrt{\frac{3 P_2 M_2^2 }{2 (M_2^2-1) }} \frac{|h_2|}{|h_1|}|h_1| s_1+\sqrt{\frac{3 P_2}{2(M_2^2-1)}}|h_2|s_2= \sqrt{\frac{3 P_2}{2(M_2^2-1)}}|h_2|(M_2 s_1+s_2)$. Recall that $s_1 \in \mathcal{A}_{M_1}=\{\pm (2k-1)\}_{k=1}^{{M_1}/2}$ and $s_2 \in\mathcal{A}_{M_2}=\{\pm (2k-1)\}_{k=1}^{{M_2}/2}$, and therefore $M_2 s_1+s_2 \in\mathcal{A}_{M_1 M_2}=\{\pm (2k-1)\}_{k=1}^{{M_1 M_2}/2}$. The quadrature component of the sum-constellation is identical to that of the in-phase component. Hence, the sum-constellation is an $M_1^2 M_2^2$-QAM constellation with a minimum Euclidean distance $d_{\rm noma}$. The case $\frac{|h_2|}{|h_1|} >\sqrt{\frac{P_1 (M_2^2-1)}{P_2 M_2^2(M_1^2-1)}}$ can be proved in a similar manner and hence is omitted for brevity.
~\hfill$\square$

\subsection{Proof of Corollary~\ref{cor:distnomavsoma}}\label{appendix:compnomaoma}
Recall that $d_{\rm noma}$ and $d_{\rm oma}$ given in~\eqref{eqn:minDist} and~\eqref{eqn:minDistoma}, respectively. We consider the following cases one by one as follows:

1) If $\frac{ M_2^2(M_1^2-1)}{M_2^2-1} \le  \frac{P_1 |h_1|^2 } {P_2 |h_2|^2}$, we have $d_{\rm noma}=\sqrt{\frac{3 P_2}{2(M_2^2-1)}}|h_2|$, and then $\frac{d_{\rm noma}}{d_{\rm oma,2}}=\sqrt{M_2^2+1}>1$.

2) If $\frac{ M_2^2 (M_1^2-1)}{ M_1^2 (M_2^2-1)} \le \frac{P_1 |h_1|^2 } {P_2 |h_2|^2} < \frac{M_2^2 (M_1^2-1)}{M_2^2-1}$, we attain $d_{\rm noma}=\sqrt{\frac{3 P_1 }{2M_2^2 (M_1^2-1)}}|h_1|$ and then we consider the following two scenarios:
		(a) For $M_2 \le M_1$, we conclude $\frac{d_{\rm noma}}{d_{\rm oma,1}}=\sqrt{\frac{M_1^2+1}{M_2^2}}>1$;
		(b) For $M_2 > M_1$, we attain $\frac{d_{\rm noma}}{d_{\rm oma,2}}= \sqrt{\frac{P_1|h_1|^2 (M_2^4-1)}{P_2 |h_2|^2 M_2^2 (M_1^2-1)}}$. As $\frac{P_1 |h_1|^2}{P_2 |h_2|^2} \ge \frac{ M_2^2 (M_1^2-1)}{M_1^2 (M_2^2-1)}$, we attain $\frac{d_{\rm noma}}{d_{\rm oma,2}}\ge \sqrt{\frac{M_2^2+1}{M_1^2}}>1$.
	
3) If $\frac{M_1^2-1}{ M_1^2 (M_2^2-1)}\le \frac{P_1 |h_1|^2} {P_2 |h_2|^2}< \frac{ M_2^2 (M_1^2-1)}{ M_1^ 2 (M_2^2-1)}$, we have $d_{\rm noma}=\sqrt{\frac{3 P_2|h_2|^2}{2 M_1^2 (M_2^2-1)}}$. Likewise, we consider the following two scenarios:
	 	(a) For $M_1 \le M_2$, then
	 	$\frac{d_{\rm noma}}{d_{\rm oma,2}}=\sqrt{\frac{M_2^2+1}{M_1^2}}>1$.
	 	(b) For $M_1>M_2$, then $\frac{d_{\rm noma}}{d_{\rm oma,1}}=\sqrt{\frac{P_2|h_2|^2 (M_1^4-1)}{P_1 |h_1|^2 M_1^2 (M_2^2-1)}}$. As $ \frac{P_1 |h_1|^2} {P_2 |h_2|^2}< \frac{ M_2^2 (M_1^2-1)}{ M_1^ 2 (M_2^2-1)}$, we have $\frac{d_{\rm noma}}{d_{\rm oma,1}}>\sqrt{\frac{M_1^2+1}{M_2^2}}>1$.
	
4) If $\frac{P_1 |h_1|^2}{P_2 |h_2|^2} <\frac{M_1^2-1}{ M_1^2 (M_2^2-1)}$, we attain $\sqrt{\frac{3 P_1}{2(M_1^2-1)}}|h_1|$, and hence $\frac{d_{\rm noma}}{d_{\rm oma,1}}=\sqrt{M_1^2+1}>1$.

From the above discussion, we can conclude that $d_{\rm noma}>d_{\rm oma}$ and this completes the proof.~\hfill$\Box$
\end{appendices}

\small
\bibliographystyle{ieeetr}

\bibliography{tzzt}

\normalsize
\end{document}